\documentclass[12pt,notoc]{JHEP3}
\usepackage{amsmath,amssymb,euscript,array,mathrsfs}
\newcommand{\startappendix}{
\setcounter{section}{0}
\renewcommand{\thesection}{\Alph{section}}}
\newcommand{\Appendix}[1]{
\refstepcounter{section}
\begin{flushleft}
{\large\bf Appendix \thesection: #1}
\end{flushleft}}
\usepackage{epsfig}
\usepackage{graphicx}

\def\Tr{{\rm Tr}}

\def\R{\boldsymbol{R}}

\def\R{R_{\rm AdS}}

\newcommand{\Dslash}{D\mkern-11.5mu/\,} %% generalized Dirac operator
\newcommand{\delslash}{{\partial\mkern-9mu/}}

\newcommand{\dd}[2]{\frac{d #1}{d #2}}

\def\Dbarslash{\,\,{\raise.15ex\hbox{/}\mkern-12mu {\bar D}}}
\def\Dslash{\,\,{\raise.15ex\hbox{/}\mkern-12mu D}}
\def\delslash{\,\,{\raise.15ex\hbox{/}\mkern-9mu \partial}}
\def\delbarslash{\,\,{\raise.15ex\hbox{/}\mkern-9mu {\bar\partial}}}

\def\a{\alpha}
\def\b{\beta}

\def\d{\delta}
\def\e{\epsilon}

\def\m{\mu}
\def\n{\nu}
\def\o{\theta}
\def\p{\pi}

\def\t{\tau}

\def\w{\omega}
\def\x{\chi}

\def\W{\Omega}
\newcommand{\EQ}[1]{\begin{equation} #1 \end{equation}}
\newcommand{\AL}[1]{\begin{subequations}\begin{align} #1
\end{align}\end{subequations}}
\newcommand{\SP}[1]{\begin{equation}\begin{split} #1
\end{split}\end{equation}}

\title{Real time response on $dS_3$:  
the Topological AdS Black Hole and the Bubble}
\author{Jimmy A. Hutasoit ${}^a$, 
S. Prem Kumar ${}^b$ and James Rafferty ${}^b$\\\\
{\it 
${}^a$ Department of Physics,\\Carnegie Mellon University,\\
Pittsburgh, PA15213, USA.
}\\
E-mail: \email{jhutasoi@andrew.cmu.edu}
\\\\
{\it 
${}^b$ Department of Physics, \\Swansea University, \\ 
Singleton Park, Swansea, SA2 8PP, U.K.
}\\
Email: \email{s.p.kumar@swansea.ac.uk},
\email{pyjames@swansea.ac.uk}
} 
\abstract{ We study real time correlators in strongly coupled ${\cal
    N}=4$ supersymmetric Yang-Mills theory on $dS_3\times S^1$, with
  antiperiodic boundary conditions for fermions on the circle. When
  the circle radius is larger than a critical value, the dual geometry
  is the so-called ``topological $AdS_5$ black hole''. Applying the
  Son-Starinets recipe in this background
  we compute retarded glueball propagators which exhibit
  an infinite set of poles yielding the quasinormal frequencies
  of the topological black hole. The imaginary parts of the propagators
  exhibit thermal effects associated with the Gibbons-Hawking
  temperature due to the cosmological horizon of the de Sitter boundary. 
We also obtain R-current correlators
  and find that after accounting for a small subtlety, the
  Son-Starinets prescription yields the retarded Green's functions. The
  correlators do not display diffusive behaviour at late times. Below
  the critical value of the circle radius, the topological black hole
  decays to the $AdS_5$ ``bubble of nothing''. Using a high frequency WKB
  approximation, we show that glueball correlators in this phase exhibit
  poles on the real axis. The tunnelling from the black
  hole to the bubble is interpreted as a hadronization transition.
} 
\begin{document}
\section{Introduction}
Time dependent backgrounds in gravity and in string theory are of
great interest from  the standpoint of the AdS/CFT correspondence
\cite{Maldacena:1997re, magoo} and related holographic dualities
between gauge theories and gravity. Time dependent classical gravity
backgrounds, in asymptotically Anti-de-Sitter spacetimes, can
potentially provide a fully nonperturbative description of
non-equilibrium phenomena in the strongly coupled dual gauge
theories. Such non-equilibrium physics in field theories arises, most
notably, in cosmology and in heavy ion
collisions at RHIC. To understand how gauge/gravity dualities work for
such processes, it is important to investigate how holography applies
in various examples with explicit time dependence. In this paper we
attempt the holographic computation of real time 
correlators of the boundary gauge theory dual to the time  
dependent, asymptotically {\em locally} AdS backgrounds found in
\cite{Birmingham:2002st, Balasubramanian:2002am, Cai:2002mr, Ross:2004cb, Balasubramanian:2005bg}.  

The authors of \cite{Aharony:2002cx} studied the double
analytic continuations of vacuum solutions such as Schwarzschild and
Kerr spacetimes providing examples of smooth, time dependent solutions
called ``bubbles of nothing'' 
\cite{Witten:1981gj, Myers:1986un, Dowker:1995gb}. These
asymptotically flat solutions were generalized to asymptotically
locally AdS spacetimes in
\cite{Birmingham:2002st,Balasubramanian:2002am}, by considering the
double analytic continuations of AdS black holes \footnote{For the classifications of solutions obtained by analytically continuing black hole solutions, see \cite{Astefanesei:2005eq}.}. The bubbles are
obtained by analytically continuing the time coordinate to Euclidean
signature $t\to i\chi$ where $\x$ is periodically identified, 
and  a polar angle $\theta \to i\t$. In addition, the $\x$ circle has
supersymmetry breaking boundary conditions for fermions. The 
resulting ``bubbles''  undergo exponential de Sitter expansion (and
contraction). For the asymptotically locally $AdS_5 \times S^5$ case
\cite{Balasubramanian:2002am}, the conformal boundary of the geometry is $dS_3
\times S^1$. The corresponding dual field theory, ${\cal N}=4$ SYM,
is thus formulated on $dS_3 \times S^1$ with antiperiodic
boundary conditions for the fermions around $S^1$. Each of the two
AdS-Schwarzschild black holes (the small and big black holes) yield an
AdS bubble of nothing solution, only one of which is stable. The
bubble of nothing geometries are vacuum solutions with cosmological
horizons \cite{Aharony:2002cx} and particle creation effects.

It was realized in \cite{Cai:2002mr, Ross:2004cb, Balasubramanian:2005bg} that
there is another spacetime with the same AdS asymptotics as the
bubble geometries, with $dS_3\times S^1$ conformal boundary. This is
the so-called ``topological black hole'' 
\footnote{The term ``topological AdS black hole'' 
has also been used to refer to black holes
with a hyperbolic horizon having a non-trivial topology. In the
AdS/CFT context these have been studied in 
\cite{Emparan:1999gf,Alsup:2008fr,Koutsoumbas:2008yq,Koutsoumbas:2008wy}
and references therein.}
-- 
a quotient of AdS space obtained by an
identification of global $AdS_5$ along a boost 
\cite{Banados:1997df, Banados:1998dc}. It is the five dimensional
analog of the BTZ black hole \cite{Banados:1992gq,
  Banados:1992wn}. The topological AdS black hole can also be obtained by a
Wick rotation of thermal AdS space. As Euclidean thermal AdS space can
be unstable to decay to the big AdS black hole via the first order Hawking-Page
transition \cite{Hawking:1982dh, Witten:1998qj}, 
a similar instability is associated to the topological AdS
black hole. In this case the topological AdS black hole is unstable to
semiclassical decay via the nucleation of an AdS bubble of nothing. The
associated bounce solution is the Euclidean small AdS-Schwarzschild
black hole which has a non-conformal negative mode. The topological
black hole becomes 
unstable only when the radius of the spatial circle becomes smaller
than a critical value (in the Euclidean thermal setup this is when the
temperature exceeds a critical value). Precisely such an instability
to decay to ``nothing'' was, of course, first noted for flat space times a
circle having antiperiodic boundary conditions for fermions
\cite{Witten:1981gj}.

The two different geometries described above are dual to two
different phases of strongly coupled, large $N$ gauge theory
formulated on $dS_3\times S^1$. As in the usual thermal interpretation
wherein the field theory lives on $S^3 \times S^1$, the two phases are
distinguished by the expectation value of the Wilson loop around the
$S^1$. In the bubble of nothing phase, the circle shrinks to zero size
in the interior of the geometry and the Wilson loop is non-zero,
indicating the spontaneous breaking of the ${\mathbb Z}_N$ symmetry of
the gauge theory. The topological black hole phase is
${\mathbb Z}_N$ invariant. Unlike the thermal situation however, the
spontaneous breaking of ${\mathbb Z}_N$ invariance is not a
deconfinement transition since the circle is a spatial direction and
not the thermal circle.

Our primary motivation in this article is to understand how the behaviour
of real time correlators in the two geometries reflects the properties
and distinguishes the two phases of the ${\cal N}=4$ theory on
$dS_3\times S^1$. Since the de Sitter boundary has its own
cosmological horizon accompanied by a Gibbons-Hawking radiation
\cite{Gibbons:1977mu}, this should also be reflected in the properties of
the boundary correlation functions. An interesting feature of both the
geometries in question is that infinity is connected, {\it i.e.} the
asymptotics is unlike the (AdS-)Schwarzschild black hole whose
asymptotics consists of two disconnected boundaries. This means 
the Schwinger-Keldysh approach in 
\cite{Herzog:2002pc} is not
directly applicable. It would be interesting to understand how to
apply that idea and also the recently proposed prescription of \cite{Skenderis:2008dh, Skenderis:2008dg} in the present context. 
Instead we simply use the Son-Starinets
prescription \cite{Son:2002sd, Policastro:2002se, Son:2007vk} to
compute real time correlators in the topological AdS black 
hole geometry, by requiring infalling boundary conditions at the
horizon of the black hole.

We find that in the topological black hole phase, retarded
 scalar glueball
correlators (homogeneous on spatial $S^2$ slices of $dS_3$) 
have a simple description in frequency space. They have an
infinite number of poles in the lower half of the complex frequency
plane. As in the case of the BTZ black hole and other well known
examples, these poles represent the black hole quasinormal
frequencies \cite{Son:2002sd}. 
The Green's functions have imaginary parts and display
features closely resembling thermal physics. These features are naturally
associated to the Gibbons-Hawking temperature due to the cosmological
horizon of de Sitter space. This suggests that the ${\cal N}=4$ theory
on $dS_3\times S^1$ is in a plasma-like or deconfined state in the
exponentially expanding universe. 

We further investigate real time correlators involving spatial spherical
harmonics of conserved R-currents to find whether they 
exhibit transport properties, {\it i.e.,} if they relax via
diffusion on the expanding spatial $S^2$ slices of $dS_3$. Applying
the Son-Starinets recipe (here we have to acount for a certain
 subtlety involving discrete normalizable mode functions in de Sitter space) 
we find that the retarded propagator of the
R-current does not appear to relax hydrodynamiccally. This is likely
 due to the ``rapid'' expansion of de Sitter space,
the expansion rate of $dS_3$ being of the same order as the
Gibbons-Hawking temperature. The real time correlators are
represented in the form of a de Sitter mode expansion, which allows to
identify a natural frequency space correlator. This latter object has
isolated poles in the lower half plane and at the origin,
and its imaginary part exhibits the features characteristic of
a thermal state.

When the spatial circle is small (relative to the radius of curvature
of $dS_3$), below a critical value, the topological black hole decays
into the  AdS bubble of nothing. In this geometry, correlation
functions are not analytically calculable. However, scalar glueball
propagators can be calculated in a WKB approximation. We show that in
this approximation, the correlation functions have an infinite set of
isolated poles on the real axis in the frequency plane. We interpret
this naturally as high mass glueball-like bound states of the field
theory. The transition from the topological black hole to the bubble
of nothing by tunelling is interpreted as a hadronization process. A
related picture of hadronization was discussed in
\cite{Horowitz:2006mr}. 

The paper is organized as follows. In Section 2, we review 
properties of the topological AdS black hole. In Section 3, we perform 
the detailed holographic computation of retarded propagators of the
spatially homogeneous scalar glueball fields. We also calculate
R-current correlation functions. Section 4 is devoted to a WKB
analysis of Green's functions in the bubble of nothing phase. We
summarize our results in Section 5. 

\section{The Topological AdS Black Hole }
The so-called topological black hole of
$\cite{Balasubramanian:2005bg, Banados:1998dc}$ 
in $AdS_5$ is an orbifold of AdS space, 
obtained by an identification of points along the orbit of a
Killing vector
\EQ
{\xi = \frac{r_\chi}{R_{\rm AdS}} \left(x_4 \partial_5 + x_5 \partial_4 \right),
}
where $r_\x$ is an arbitrary real number 
and the AdS space is described as the universal covering of the hyper-surface
\EQ
{-x_0^2 + x_1^2 + x_2^2 + x_3^2 + x_4^2 - x_5^2 = - R^2_{\rm Ads},
}
$R_{\rm AdS}$ being the AdS radius.
In Kruskal-like coordinates which cover the whole spacetime,
the metric has the form
\EQ
{
ds^2= {4R^2_{\rm AdS}\over (1-y^2)^2}\;dy^\mu dy^\nu \eta_{\mu\nu}+ 
{(1+y^2)^2\over (1-y^2)^2}\;r_\x^2d\chi^2
\label{kruskal}
}
where $\chi$ is a periodic coordinate with period $2\pi$. The four
coordinates $y^\mu$, $(\mu=0,\ldots 3)$ are non-compact with the 
Lorentzian norm $y^2 = y^\mu y^\nu\eta_{\mu\nu}$ such that $-1<y^2<1$. 
Locally, the spacetime is anti-de Sitter 
with a periodic identification of the $\chi$ coordinate,
\EQ
{
\chi\sim \chi+2\pi.
}
The conformal boundary of the spacetime is approached as
$y^2\rightarrow 1$, and it is $dS_3 \times S^1$. The boundary
conformal field theory is therefore formulated 
on a three dimensional de Sitter space with radius of curvature $R_{\rm AdS}$
times a spatial circle of radius $r_\x$. 
%Defining a spatial radial
%coordinate
%\EQ
%{
%Y^2 = y_i y_i,
%}
%we obtain the global spacetime structure in Figure 1.

The geometry has a horizon at $y^2=0$, which is the three dimensional
hypercone,  
\EQ
{y_0^2= y_1^2+y_2^2+y_3^2,}
and a 
singularity at $y^2=-1$. The singularity appears because
the region where the Killing vector has
negative norm needs to be excised from the physical spacetime to
eliminate closed timelike curves. The hyperboloid $y^2=-1$ is a
singularity since timelike geodesics end there
and the Killing vector $\partial_\chi$
 generating the orbifold identification has vanishing norm at $y^2=-1$. 
%The space transverse to the singularity, in its vicinity, 
%is of the form $R^{1,1}/{\mathbb Z}$. 
The topology of the spacetime is ${\mathbb R}^{3,1}\times S^1$, in
contrast to that of the  AdS-Schwarzschild black hole which has the topology 
${\mathbb R}^{1,1}\times S^3$. For this reason, infinity is connected in this
geometry unlike in the usual Schwarzschild black hole which has two
disconnected asymptotic regions.

\begin{figure}[h]
\begin{center}
\epsfig{file=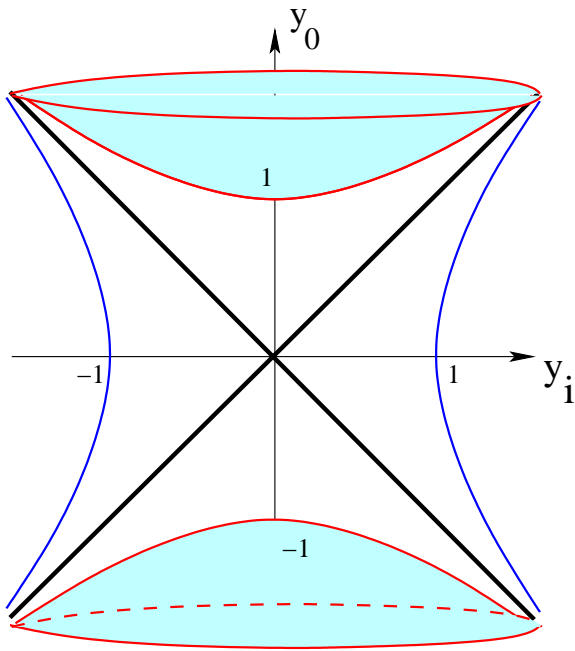,width=3.0in}
\end{center}
\caption{\footnotesize The global structure of the topological
AdS black hole spacetime. The singularity is the hyperboloid
$-y_0^2+y_i y_i=-1$ and the horizon is at the cone 
$y_0^2 = y_i y_i$.}
\label{spacetime}
\end{figure}

Finally, it is possible to rewrite the metric in Schwarzschild-like
coordinates by introducing
\EQ
{
Y^2=\sum_{i=1}^3y_iy_i\,;\qquad {Y\over y_0}=
\coth\left({t\over R_{\rm AdS}}\right)\,;
\qquad {r^2\over R^2_{\rm AdS}} = 4 {(Y^2-y_0^2)\over (1+y_0^2-Y^2)^2}
}
These coordinates only  cover the region $y^2\geq 0$ which is the
exterior of the topological black hole. Locally, the metric takes the
form (simply related to 
Eq.(11) of \cite{Balasubramanian:2005bg} after a coordinate transformation),
\SP
{
&ds^2= \\
&R^2_{\rm AdS} {dr^2\over (r^2+R^2_{\rm AdS})} + 
\left({r_\x\over R_{\rm AdS}}\right)^2
(r^2+R^2_{\rm AdS})d\chi^2+
r^2
\left(-{dt^2\over \R^2}+\cosh^2 \left({t\over R_{\rm AdS}}\right)
  \;d\Omega_2^2\right). 
\label{schw}
}
The Euclidean continuation of the metric yields thermal AdS space due
to periodicity of the $\chi$ coordinate. Hence, the topological black
hole metric (exterior to the  
horizon) can also be obtained following a double Wick rotation
of global AdS spacetime and a periodic identification of the  
$\chi$ coordinate. In the Schwarzschild-like coordinates, 
the horizon of the topological black hole is at $r=0$.
It is clear that each slice of constant $r$ is a $dS_3\times S^1$ geometry.
This metric,  while locally
describing AdS space, differs from it globally due to the
identification ${\chi\sim\chi + 2\pi}$. 
Note also that the spatial $S^1$ remains finite sized at the horizon,
with radius  $R_{S^1}= r_\x$.

It is interesting to see that 
we can get a better understanding of the geometry in the vicinity of
the singularity at $y^2=-1$ by zooming in on the 
the metric \eqref{kruskal} in this region.
Introducing the coordinates
\EQ
{y_0= (1-\delta) \, \cosh\varepsilon\,,\qquad Y =(1-\delta)\,\sinh\varepsilon
\qquad0< \delta\ll 1,}
we find
\EQ
{
ds^2\approx R^2_{\rm AdS}(-d\delta^2 + d\varepsilon^2 + \sinh^2\varepsilon
\,d\Omega_2^2)+ r_\x^2 \delta^2\,d\chi^2,
}
which is the metric for Milne spacetime (in the $\delta, \chi$ directions).

\section{Real time correlators in the Topological AdS Black Hole}

We will compute real time correlators in the 
Yang-Mills theory on the boundary of the
topological $AdS_5$ black hole (TBH) following the  
recipe of Son and Starinets \cite{Son:2002sd} in the Schwarzschild-like patch
\eqref{schw} of the black hole. 
%\footnote{Unlike the Schwarzschild-AdS black hole, as the geometry does not have two disconnected asymptotic regions, the Schwinger-Keldysh closed time path picture is not manifest in the bulk. It would be interesting to understand how to apply the procedure of \cite{Herzog:2002pc} to obtain real time correlators in this background.}
Viewing the topological black hole as 
a Wick rotation of (Euclidean) thermal AdS space, one expects that 
such correlators can also
be obtained by an appropriate analytic continuation of 
Euclidean Yang-Mills correlators on $S^3 \times S^1$ 
in the confined phase (the ${\mathbb Z}_N$ symmetric phase) with
anti-periodic boundary conditions for fermions. 
Since the relevant Wick rotation  
turns the polar angle on $S^3$ into the time
coordinate of de Sitter space, a complete knowledge of the angular
dependence of Euclidean correlators on $S^3\times S^1$would be necessary.
However, finite temperature Yang-Mills correlators on $S^3$
 and at strong coupling, 
have not been calculated explicitly, so we will not follow the route of
analytic continuation. Instead we will directly calculate the real
time correlators  
using the holographic prescription of Son and Starinets 
applied to the topological AdS black hole geometry. 

\subsection{Scalar wave equation in the Topological Black Hole}
To extract field theory correlators, we first need to look for solutions to the
wave equation in the region exterior to the horizon of the
topological black hole. 
It is instructive to write the metric for the black hole 
in the Schwarzschild form of \cite{Balasubramanian:2005bg}
\EQ
{
ds^2= \R^2 \left[{d\rho^2\over (\rho^2 -1)} + 
\left({r_\x\over \R}\right)^2
\rho^2d\chi^2+
(\rho^2-1)
\left(-d\tau^2+\cosh^2 \tau \;d\Omega_2^2\right)\right],
\label{schw1}
}
where we have introduced the dimensionless variables
\EQ
{
\rho=\sqrt{\left(r/\R\right)^2+1},\qquad\qquad\tau={t\over\R}.
} 
The conformal boundary of the space is approached as
$\rho\rightarrow\infty$ while  
the horizon is at $\rho=1$, where the coefficient of $d\tau^2$
vanishes. The slices with constant $\rho$ are manifestly $dS_3\times
S^1$. 

The scalar fields in this geometry have a natural expansion 
in terms of harmonics on the $S^2\times S^1$ spatial slices 
\EQ
{
\Phi(\rho,\chi,\tau,\Omega)
=\sum_{\ell,m,n}\;{\cal } 
A_{\ell \,m}\;Y_{\ell \,m}(\Omega)\;e^{in\chi}\int {d\nu\over
  2\pi}\;\Phi_{n}\,(\nu, \rho)\;{\cal T}_{\ell}(\nu,\tau).
\label{decomp}
}
The normal mode expansion above involves spherical harmonics on $S^2$,
the discrete Fourier modes on $S^1$ and ${\cal T}_{\ell}(\nu,\tau)$
which  solve the scalar wave equation on
$dS_3$:
\EQ
{
{1\over \cosh^2 \tau}\partial_\tau \left(\cosh^2\tau \;\partial_\tau{\cal
    T}_{\ell}(\nu,\tau)\right)+{\ell(\ell+1)\over \cosh^2\tau}{\cal T}_{\ell}(\nu,\tau) 
= -(\nu^2+1){\cal
    T}_{\ell}(\nu,\tau).
}\\
For every $\ell \in {\mathbb Z}$, the equation has two kinds of
solutions that will be relevant for us: 
i) normalizable modes labelled by integers
$- i\nu = 1,2,\ldots \ell$; and ii) delta-function normalizable modes
labelled by a continuous frequency variable $\nu \in {\mathbb R}$. We
will return to this point when we discuss R-current correlators.
General solutions to this equation can be expressed in terms of
associated Legendre functions
\EQ
{
{\cal T}_{\ell}(\nu,\tau)=\;{1\over \cosh\tau}
\left(\;A_l\;P_\ell^{i\nu}(\tanh\tau)+\;B_l\;Q_\ell^{i\nu}(\tanh\tau)\right).
}
In the usual approach to quantizing free scalar fields in de Sitter
space, the integration constants $A_\ell$ and $B_\ell$ are determined by the 
choice of de Sitter vacuum \cite{mottola, bd, Spradlin:2001pw}. However, in the present
context, the constants will be specified by picking out infalling wave
solutions at the horizon of the topological black hole. These are the
holographic boundary conditions relevant for real time response
functions in the strongly coupled field theory on $dS_3\times S^1$.

It is useful to see 
the scalar wave equation in this background recast as a
Schr\"odinger equation, using Regge-Wheeler type variables  
\EQ
{
u={1\over 2}\ln\left({\rho+1\over \rho-1}\right)\qquad{\rm or} \qquad
\rho=\coth u,
\label{rg1}
}
and 
\EQ
{
\Psi_n =  \sqrt{\rho(\rho^2-1)}\;\Phi_n,
\label{rg2}
}
$\Phi$ being the scalar field in the bulk. 
\begin{figure}[h]
\begin{center}
\epsfig{file=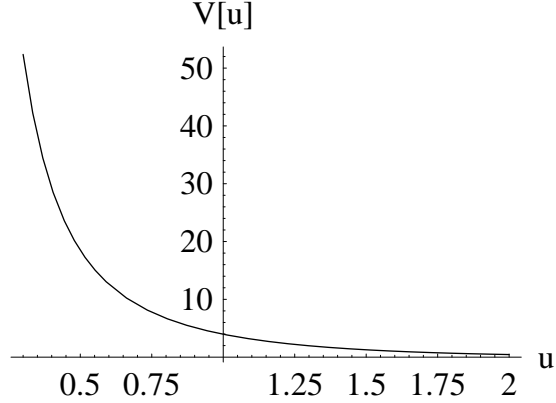,width=3.5in}
\end{center}
\caption{\footnotesize The Schr\"odinger potential for the
topological AdS black hole.}
\label{vofu}
\end{figure}
In these coordinates, the horizon is approached as $u\rightarrow
\infty$ while the conformal boundary is at $u=0$.
Following the above coordinate and field redefinitions, the
Schr\"odinger wave equation in the TBH geometry is 
\AL
{
&-{d^2\over du^2}\;\Psi_{n}(\nu,u) 
+ V_{n}(u)\;\Psi_n(\nu, u)= \nu^2\;\Psi_n(\nu,u),
\\\nonumber\\ 
&V_{n}(u)= ((M R_{\rm AdS})^2+\tfrac{15}{4}){1\over \sinh^2 u}+(\tfrac{1}{4}+
\tfrac{R^2_{\rm AdS}}{r_\x^2}\,n^2){1\over\cosh^2 u}\;. 
\label{schrodinger1}
}
Here, $\Psi_n = \Phi_n\sqrt{\rho(\rho^2-1)}$ and 
we have allowed for a generic non-zero mass $M$ since we will
eventually be interested both in the massless and massive cases.
As expected for AdS black holes, the potential decays exponentially near
the horizon $u\rightarrow \infty$, while blowing up near the boundary
at $u\rightarrow 0$. In the near horizon region where the potential
vanishes, the solutions with $\nu>0$ 
are travelling waves and there is a natural choice
of incoming and outgoing plane wave solutions. 
For any $n$ and $M$, the equation has analytically tractable 
solutions in terms of
hypergeometric functions. We will use these to calculate the retarded
Green's functions for the boundary gauge theory ({\it i.e.,} ${\cal N}=4$
SYM) at strong coupling on $dS_3\times S^1$.

Although analytical solutions exist for all non-zero $n$ and
$M$, we will restrict attention, for simplicity, to two special cases:
i) $n\neq 0$ and 
$M R_{\rm AdS}=0$; 
ii) $n=0$ and $M R_{\rm AdS}\neq 0$. In each of these two cases 
the radial equation is solved by two linearly independent
hypergeometric functions:
\SP
{&\underline {M=0;\quad n\neq 0}: \\\\
&\Phi_n(\nu,\rho)\,=\,C_1\; \rho^{-i\,n\,\R/r_\x}\,\,\times\\
&\qquad(\rho^2-1)^{-i{\nu\over 2}-{1\over 2}}\;
{}_2
F_1\left(-\tfrac{1}{2}-\tfrac{i}{2}(\nu+n\,\tfrac{\R}{r_\x}),
\;\tfrac{3}{2}-\tfrac{i}{2}(\nu+n\,\tfrac{\R}{r_\x});
\;1-i n\,\tfrac{\R}{r_\x};\;\rho^2\right)\\
&+\,C_2 \; \rho^{i \,n\,\R/r_\x}\,\,\times\\
&\qquad\;(\rho^2-1)^{-i{\nu\over 2}-{1\over 2}}\;
{}_2
F_1\left(-\tfrac{1}{2}-\tfrac{i}{2}(\nu-n\,\tfrac{\R}{r_\x}),
\;\tfrac{3}{2}-\tfrac{i}{2}(\nu-n\,\tfrac{\R}{r_\x});
\;1+i n \,\tfrac{\R}{r_\x};\;\rho^2\right).
%&p_+=\tfrac{\R}{r_\x}.
\label{massless}
}
Similarly, for massive fields with $n=0$, the two linearly independent
solutions are
\SP
{
&\underline{M \neq0;\quad n=0}:\\\\
&\Phi_0(\nu,\rho)=\\
&\;C_1\;
(\rho^2-1)^{{-\tfrac{1}{2}}(4-\Delta)}\; 
{}_2F_1\left(\tfrac{1}{2}(3-\Delta)-\tfrac{1}{2}i\nu,\;    
\tfrac{1}{2}(3-\Delta)+\tfrac{1}{2}i   
\nu;\;3-\Delta;\;-(\rho^2-1)^{-1}\right)\\\\
& +\,C_2\; (\rho^2-1)^{-\tfrac{1}{2}\Delta}\;
{}_2F_1\left(\tfrac{1}{2}(\Delta-1)-
\tfrac{1}{2}{i \nu},
\;\tfrac{1}{2}(\Delta-1)+\tfrac{1}{2}i{\nu};\;\Delta-1;\;
-(\rho^2-1)^{-1}\right)\\\\
&\Delta=2+\sqrt{4+(M\R)^2}.
}
The correct linear combination, relevant for the holographic
computation of correlators, is  picked by applying the requirement
of purely infalling waves at the horizon of the topological black hole
at $\rho\approx1$.

\subsection{Scalar glueball correlator}

As argued in \cite{Balasubramanian:2002am, Balasubramanian:2005bg}, 
the topological black hole
in AdS space is automatically a solution to the Type IIB supergravity
equations of motion, since it can be obtained via a double Wick rotation
(and an identification) of $AdS_5\times S^5$. 
The ${\cal N}=4$ supersymmetric Yang-Mills theory on the $dS_3\times S^1$
conformal boundary of the topological black hole, has two $SO(6)$
singlet, scalar glueball fields 
\EQ
{
{\cal G}(\vec x,t)=\Tr F_{\mu\nu}F^{\mu\nu};\qquad
\tilde {\cal G}(\vec x,t)=\Tr F_{\mu\nu}\tilde F^{\mu\nu}.
}
These are dual to the dilaton and the RR-scalar in the Type IIB theory
on the bulk spacetime and 
 both solve the massless (corresponding to $\Delta=4$ operators on the
 boundary) scalar wave equation in the topological
 black hole geometry. The retarded propagators for the scalar glueball
 fields are known on ${\mathbb R}^{3,1}$ at weak coupling both at zero
 and finite temperature \cite{Hartnoll:2005ju}. 
Since the operator is chiral primary in the
 ${\cal N}=4$ theory, at zero temperature its propagator on ${\mathbb
   R}^{3,1}$ receives no quantum corrections and the strong coupling
 results from supergravity are in exact agreement with those of
 the free field theory. At finite temperature, however, when
 supersymmetry is broken, strong and
 weak coupling results on ${\mathbb R}^{3,1}$ differ \cite{Son:2002sd,
   Hartnoll:2005ju}. Computations of the glueball correlator also
 exist in the
 free ${\cal N}=4$ theory at finite temperature and on a spatial $S^3$, both
 in the confined and deconfined phases \cite{Hartnoll:2005ju}. Their
 strong coupling counterparts have not been determined.

The present case, with the field theory on $dS_3 \times S^1$, 
is intriguing for the following reasons. First, there is the lack of
supersymmetry, 
due to antiperiodic boundary conditions for fermions around the
spatial $S^1$. Secondly, the
boundary field theory sees a cosmological horizon on $dS_3$ 
accompanied by its
thermal bath. It would be interesting to observe the emergence of
the boundary Gibbons-Hawking temperature  
from a holographic calculation of its
correlators at strong coupling. Finally, when the radius of the boundary $S^1$
decreases below a critical value, $r_\x/\R \leq {1\over 2\sqrt{2}}$, the
topological black hole decays via a bounce to the small ``AdS bubble of
nothing''. We would like to understand how 
boundary field theory correlators at strong coupling 
on $dS_3\times S^1$ change across this transition. The 
transition from the topological black hole to the Bubble of Nothing is
a ${\mathbb Z}_N$ breaking transition. This is understood precisely as
in the Euclidean (finite temperature) situation, due to a non-zero
expectation value for the Wilson loop around the spatial $S^1$.

Curiously, it is apparent that in the classical supergravity
approximation, the bulk scalar glueball correlators are insensitive to 
fermions and their boundary conditions around the spatial $S^1$. 
It would be interesting to understand whether 
this is related to large-$N$ volume independence 
\cite{Kovtun:2007py,Unsal} in the ${\mathbb Z}_N$ symmetric phase.

We are primarily interested in real time response and for
the sake of simplicity, we will first 
study only the response functions for glueball fluctuations
that are homogeneous on the spatial $S^2$ slices at the boundary, {\it
  i.e.},
\SP
{
%\langle {\cal G}^{0}_n(t){\cal G}^{0}_n(0)\rangle_R
&G_R(\t,\t'\,;n\,;l=0)
= \\
&-i\int \frac{d\Omega}{4\pi}\,\frac{d\Omega'}{4\pi} 
\;\int{d\chi\over 2\pi} \;e^{-in \chi} 
\,\Theta(\t-\t')\;\langle\,\left[\;{\cal G}(\Omega,\chi,\t),\;{\cal
    G}(\Omega',0,\t')\;\right]\,\rangle.  
}
We will work with the dimensionless variables $\t= t/R_{\rm AdS}$,  
$\t' = t'/R_{\rm AdS}$ and restore appropriate dimensions when necessary. 
As we will see when we look at R-current correlators, it is 
straightforward to generalize to 
inhomogeoneous fluctuations on the spatial sphere. 

For the moment we focus attention on the $s$-wave 
($\ell=0$) retarded correlation function of the
scalar glueball operator. Also, for the $s$-waves, the correlator turns out to
be a function of $(\t-\t')$ so that it is natural to 
define the temporal Fourier transform of this as,
\EQ
{
\tilde G_R(\nu\,;n) = \int _{-\infty}^\infty {d\t}
\, e^{-i\nu (\t-\t')} \, G_R(\t,\t'\,;n\,;\ell=0).
}
To calculate it at strong coupling and in the large radius regime
($r_\x\geq {\R\over2\sqrt 2}$), 
we solve the dilaton wave equation
which is the equation for a massless, minimally coupled scalar field in the
background of the topological AdS black hole. 

\subsubsection{Spatially homogeneous case with $n=\ell=0$ }
We begin by looking at the spatially
homogeneous response functions, $l=n=0$, 
on the $dS_3\times S^1$ slices. The solutions to
the radial part of the Klein-Gordon equation in the massless limit 
are the hypergeometric functions
\EQ
{
{\Phi}_0^{(1)}(\nu,\rho)=\frac{\pi}{4}\;{1+\nu^2\over\cosh
\left(\tfrac{\pi \nu}{2}\right)}
\;\;{}_2F_1\left(\;-\tfrac{1+i\nu}{2}\;,\;-\tfrac{1-i\nu}{2}\;;\;1\;;
  \;\tfrac{\rho^2}{\rho^2-1}\right)     
%\Gamma\left(\tfrac{3 -i s}{2}\right)\Gamma\left(\tfrac{3+is}{2}\right)
}
and
\EQ
{
{\Phi}_0^{(2)}
(\nu,\rho)=(\rho^2-1)^{-2}\;{}_2F_1\left(\tfrac{3-i\nu}{2}\;,\;
  \tfrac{3+i\nu}{2}\; 
;\;3\;; \;-\tfrac{1}{\rho^2-1}\right).
}
For the $l=0$ modes, the temporal dependence is also particularly
simple, and has a natural interpretation in terms of positive and
negative frequency states
\EQ
{
{\cal T}_{0}^+(\nu,\tau)= {e^{-i \nu \tau}\over \cosh\tau} \qquad
{\rm and}\qquad
{\cal T}_{0}^-(\nu,\tau) ={e^{i \nu \tau}\over \cosh\tau}.
\label{modes}
}
%\EQ
%{
%z^5\partial_z\left({1+z^2\over z^3}\partial_z\phi\right) -
%{z^2\over \cosh^2 t}\partial_t \left(\cosh^2t \;\partial_t\phi\right)
%- m^2\phi=0.
%}
Solving the Dirichlet problem and extracting correlation
functions holographically requires us to
first pick the correct linear combination of the
two solutions which is smooth near the horizon
$\rho\rightarrow 1$ and represents an incoming wave falling
into the horizon. 
In the near horizon region, the asymptotic form of the solutions is:
\SP
{
&{\Phi}^{(1)}_0(\nu, \rho\rightarrow1) \rightarrow\\ 
%\frac{\pi}{4}\;{1+\nu^2\over\cosh 
%\left(\tfrac{\pi \nu}{2}\right)} \times \notag \\ & \times 
%\left(
&i\;(2(\rho-1))^{-\frac{1-i \nu}{2}} \;e^{{\pi\over 2}\nu}\;
{\Gamma(-i \nu)\;\Gamma\left({3+i\nu\over2}\right)\over 
\Gamma\left(-{1+i\nu\over 2}\right)}
+i\;(2(\rho-1))^{-\frac{1+i\;\nu}{2}}
\;e^{-{\pi\over 2}
\nu}{\Gamma(i \nu)\;\Gamma\left({3-i\nu\over2}\right)
\over \Gamma\left(-{1-i\nu\over 2}\right)} 
%\right)
}
and
\EQ
{
{\Phi}^{(2)}_0(\nu, \rho\rightarrow1)\rightarrow 
 \;(2(\rho-1))^{-\tfrac{1-i \;\nu}{2}} 
{2\; \Gamma(-i \nu)\over 
\Gamma\left({3-i\nu\over 2}\right)^2}
+\;(2(\rho-1))^{-\tfrac{1+i\;\nu}{2}} {2 \;\Gamma(i
\nu) 
\over \Gamma\left({3 + i\nu\over 2}\right)^2}.
}

Note that these modes diverge as $1/\sqrt{(\rho-1)}$ near the
horizon. However, employing the measure implied by the bulk metric
$\sqrt {-g} \sim (\rho^2-1)^{5/2}$, these are still normalizable in the vicinity
of the horizon. We can now 
pick a linear combination such that only incoming positive frequency 
waves are allowed at the
horizon. 
This means, assuming that ${\cal T}_{0}^+$ are the 
positive frequency modes with ${\rm
  Re}(\nu)>0$, 
the solution to the radial wave equation should behave like
$(\rho-1)^{-\frac{1+i\nu}{2}}$ near the horizon. 
In conjunction with this we have the properly normalized 
boundary behaviour as $\rho \rightarrow \infty$,
\EQ
{
\Phi^{(1)}_0(\nu,\rho\rightarrow \infty)\rightarrow 1 +\ldots; \qquad 
{\Phi}^{(2)}_0(\nu, \rho\rightarrow \infty)\rightarrow
{1\over\rho^4}
+ \ldots.}
The complete solution to the boundary value problem for a massless
scalar with $l=n=0$, in the topological AdS black hole is then,
\EQ
{
{\Phi}_0(\nu,\rho)= {\Phi}^{(1)}_0(\nu,\rho) +
i {\pi\over 32}\;e^{{\pi\over
    2}\nu}\;{(\nu^2+1)^2\over\cosh{{\pi\over 2}\nu}}
\;{\Phi}^{(2)}_0(\nu,\rho).
\label{sol}
}
Following the holographic 
prescription \cite{Son:2002sd,Policastro:2002se} for computing real
time correlators, the Yang-Mills 
retarded correlation function is obtained by analyzing the
boundary terms from the on-shell scalar action
\SP
{
S
&= {N^2\over 16\pi^2}
\int d\tau\int d\Omega\int d\chi\;g^{\rho\rho}
\sqrt{-g}\; \;\Phi(\tau,\rho)\;\partial_\rho
\Phi(\tau,\rho)\;\big|_{\rho\rightarrow \infty}
\\
}
where, for the spatial $s$-wave we have defined
\EQ
{
\Phi(\tau,\rho)= \int_{-\infty}^{\infty} {d\nu\over 2\pi} \; {\cal
  T}_{0}^+(\nu,\tau)\;\Phi_0(\nu,\rho).
} 
%{
%&\hspace{3in}\times{\cal Z}_{0,0}(z,\omega)
%\partial_z{\cal Z}_{0,0}(z,\omega')\big|_{z\rightarrow 0}
%.}
Putting together the explicit expressions for ${\cal
  T}_{0}^+(\nu,\tau)$, and the 
boundary behaviour of the solution \eqref{sol} we are immediately led
to the (unrenormalized)
$s$-wave retarded correlator in frequency space, including all contact
terms (finite polynomials in the frequency $\nu$)
\SP
{
\tilde G_R(\nu; 0)= &{N^2\over 16\pi^2}\left(-\tfrac{1}{8}(1+\nu^2)^2
\left[\psi\left({3-i\nu\over2}\right)+ 
\psi\left({3+i\nu\over2}\right)- i\pi
\coth\left(\pi\tfrac{1}{2}(\nu+i)\right)
\right]\right.\\
&\left.+\tfrac{1}{4}(1+\nu^2)^2\;\left(\ln \rho
  -\gamma_E+1\right)\big|_{\rho \rightarrow \infty}+ 
\tfrac{1}{2}(1+\nu^2){\rho^2}\big|_{\rho\rightarrow \infty}\right).
\label{ret}
}
We remark that unlike the case of the Poincare' patch description of 
AdS space, the
non-normalizable solution in the topological AdS black hole (akin to
global AdS), contains a term proportional to $1/\rho^2$ in its near-boundary
asymptotics, but it only
contributes a quadratically divergent contact term in the correlation function
above. 
\begin{figure}[h]
\begin{center}
\epsfig{file=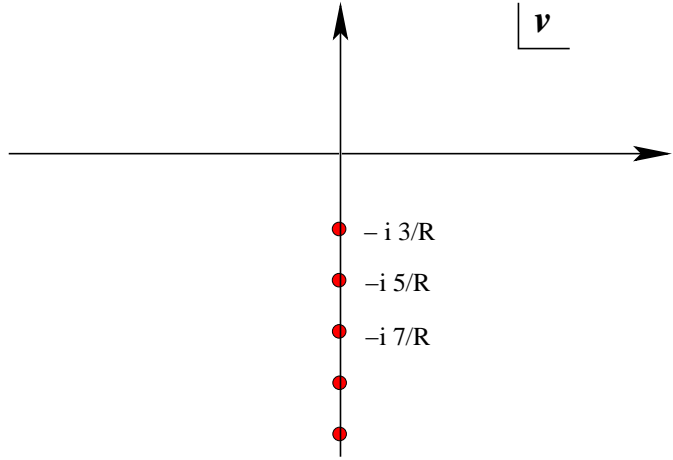,width=3.5in}
\end{center}
\caption{\footnotesize The analytic structure in complex
frequency plane, of the massless,
spatially homogeneous ($l=n=0$), retarded Green's function in the
${\mathbb Z}_N$ symmetric phase, corresponding to the topological
AdS black hole.}  
\label{retbon}
\end{figure}

The divergent and scheme-dependent contact terms can be minimally
subtracted away to yield the renormalized, retarded Green's function. 
Up to now we have been working with dimensionless variables,
corresponding to a de Sitter boundary of unit curvature. Restoring
dimensionful constants with the replacement
\EQ
{
\nu \rightarrow
\nu R,
}
$R$ being the radius of curvature of the boundary $dS_3$ (or the
inverse Hubble constant), the renormalized retarded Green's function
continued into the complex frequency plane is 
\EQ
{
\tilde G_R(\nu;0)=-\frac{N^2}{64\pi^2}(R^{-2}+\nu^2)^2 \;
\left[
\psi\left({3 -i\nu R\over2}\right)
-{2i\nu R \over(1+\nu^2 R^2)}\right].
\label{homog}
}
For ${\rm Im}(\nu)=0$, its real and imaginary parts match
\eqref{ret}, and the function is analytic in the upper half plane,
with only isolated simple poles in the lower half plane at
\EQ
{
\nu_k = -i(3+ 2k){1\over R}\qquad k\in {\mathbb Z}
}
As argued in \cite{Son:2002sd}, poles of the retarded correlator
in a black hole background coincide with the quasinormal
frequencies of the black hole. The quasinormal frequencies
and the retarded glueball 
correlator found above, 
for the topological black hole in $AdS_5$, are remarkably
similar to the corresponding objects in the BTZ black hole
\cite{Son:2002sd}.
\subsubsection{Non-zero momentum along $S^1$ and $l=0$}

It is relatively easy to allow for a non-zero discrete momentum $n/r_\x$
along the spatial $S^1$. This requires
the modes \eqref{massless}, to solve the
boundary value problem in the topological black hole
background. Following the same steps as in the s-wave correlator,
(after tedious algebra)
we find that the retarded Green's function (with
dimensionful constants restored) is
\SP
{
&\tilde G_R(\nu\,; n)=
-\frac{N^2}{128\pi^2}\left((\nu
  -\tfrac{n}{r_\x})^2+R^{-2}\right) 
\left((\nu +\tfrac{n}{r_\x})^2+ R^{-2}\right)\, \times\\
&\left[\psi\left(\frac{3}{2}- i\tfrac{R}{2}(\nu  - 
    \tfrac{n}{r_\x})\right)+\psi\left(\frac{3}{2}-
    i\tfrac{R}{2}(\nu  + \tfrac{n}{r_\x})\right)
- {2iR(\nu - \tfrac{n}{r_\x})
\over(\nu - \tfrac{n}{r_\x})^2R^2+1}\right.\\
&\left.-{2i R(\nu +\tfrac{n}{r_\x})
\over(\nu +\tfrac{n}{r_\x})^2R^2+1}\right]\,,\qquad
n\in {\mathbb Z}.
}
When $n=0$, this matches our expression for the $s$-wave correlator
\eqref{homog}.
The Green's function has nonanalyticities only in the lower half
plane, with simple poles at  
\EQ
{
\nu_k^\pm = - i(3+2k)R^{-1} \pm {n\over r_\x};\qquad k,n\in {\mathbb
  Z},
}
giving the quasinormal frequencies of the topological black hole, with
non-zero momentum along the spatial $S^1$.
Interestingly each simple pole at $n=0$ `splits' into 
two simple poles at non-zero $n$.
\begin{figure}[h]
\begin{center}
\epsfig{file=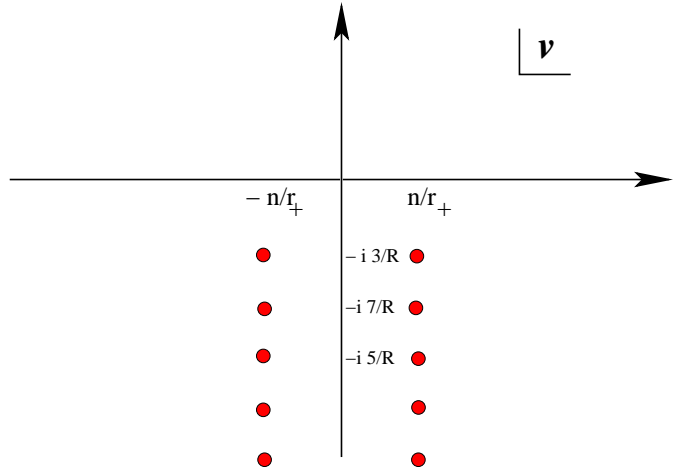,width=3.5in} 
\end{center}
\caption{\footnotesize Simple poles in the
frequency plane, of the massless retarded Green's function with non-zero 
momentum along the spatial circle.}
\label{retdouble}
\end{figure}
Our expressions for the correlation functions on $dS_3\times S^1$ in
the ${\mathbb Z}_N$ symmetric phase satisfy the obvious
consistency check -- in the high frequency/large momentum limit $\nu R, n
\gg 1$, they reproduce the flat space Green's function for the scalar glueball
operator
\EQ
{
\tilde G_R(\nu\,;n)\Big|_{\nu R,\,n\gg1}\longrightarrow -{N^2\over
  128\pi^2}(\nu^2-\tfrac{n^2}{r_\x^2} )^2\;\ln(\nu^2-\tfrac{n^2}{r_\x^2} )
.}
\subsection{Thermal effects and the Gibbons-Hawking temperature}
De Sitter space has a cosmological horizon and an associated Gibbons-Hawking
temperature \cite{Gibbons:1977mu}. We therefore expect our boundary
($dS_3\times S^1$) field theory correlators to exhibit thermal
properties. For real frequencies $\nu$, the digamma functions have
an imaginary part so that
\SP
{
{\rm Im} \;\tilde G_R(\nu\,;n)\Big|_{{\rm Im}(\nu)=0}=&
{N^2\over 256\pi}\left((\nu
  -\tfrac{n}{r_\x})^2+R^{-2}\right) 
\left((\nu +\tfrac{n}{r_\x})^2+ R^{-2}\right)\times\\
&\left[\coth\left(\frac{\pi R}{2}(\nu+\tfrac{n}{r_\x}-iR^{-1})\right)
+\coth\left(\frac{\pi R}{2}(\nu+\tfrac{n}{r_\x}-iR^{-1})\right)
\right]
\label{imds}
}
Here we have used $\tanh(x)=\coth(x+i\tfrac{\pi}{2})$, to cast the
result in a form that will make the connection to thermal physics
explicit. 

In {\em flat space} and in free field theory at finite temperature
$T\neq 0$, the scalar glueball 
propagator , with zero spatial momentum and frequency $\omega$
is proportional to the digamma
function \cite{Hartnoll:2005ju}
\EQ
{
\tilde G_R(\omega)\big|_{\rm Flat \;space}=-{N^2\over 2\pi^2
}\;\omega^4 \psi\left(- i 
\omega\over 4\pi T\right) + {\rm analytic}.
}
The imaginary part of the flat space glueball correlator is the
spectral function
\EQ
{
{\rm Im}\;\tilde G_R(\omega)\big|_{\rm Flat\;space}= -{N^2\over 2\pi^2}
\omega^4 \pi \coth\left( {\omega\over 4 T}\right).
\label{imflat}
}
In weakly coupled, or free field theories at finite temperature (on
flat space), the one-loop spectral function reflects the physical effect of 
`Bose enhancement', following from stimulated emission of bosons into
the heat bath. In perturbative field theory on flat space, this
manifests itself as an enhancement of the decay rate
 of an unstable boson in a heat bath, 
at rest with energy $\omega$,   by a factor (relative to the vacuum
decay rate)
\EQ
{
\coth\left({\omega\over 4 T}\right)= 1+ 2n_B\left(\tfrac{\omega}{2}\right) = 
1+{2\over e^{\omega/2T}-1}.
}
While there may not be an obvious way define a spectral representation in de
Sitter space, the similarity between our strongly coupled de Sitter
space result \eqref{imds} and \eqref{imflat}
is obvious. In particular, it allows the identification of a temperature in 
de Sitter space
\EQ
{
T_H = {R^{-1}\over 2 \pi} 
}
which is precisely the value of the Gibbons-Hawking temperature. Note
that despite the similarity between the expressions for $dS_3$ and
thermal correlators in 
flat space, there is a crucial difference -- the frequency or
`energy' appearing in the Bose-Einstein-like 
distribution function in de Sitter
space, is not the real frequency $\nu$ 
\eqref{imds}, but in fact $\nu - i R^{-1}$. 
This difference can be traced back to
the definition of our positive and negative frequency modes
\eqref{modes}. For real $\nu$, the positive frequency modes are red-shifted
away in the far future. To get propagating modes in the future, we
would need to choose $\nu =\omega+ i R^{-1}$ with $\omega\in {\mathbb
  R}$.

It is interesting to note that our results for the retarded
glueball propagator in the strongly coupled field theory (in the
${\mathbb Z}_N$ symmetric phase) on de Sitter space closely match one-loop
weakly coupled field theory calculations \cite{Boyanovsky:1996ab}. It would be a
straightforward calculation to check whether there is exact agreement
between weak and strong coupling results on $dS_3\times S^1$. We leave
this excercise for the future. For now, we only make the following
observation, which suggests that the scalar glueball correlator in
the ${\mathbb Z}_N$ symmetric phase should not be renormalized.

It has been argued in \cite{Kovtun:2007py, Unsal} that
correlation functions of large $N$ gauge theories in the 
${\mathbb Z}_N$ symmetric phase, with some or all spacetime
directions compactified, should be independent of the volume of the
compact directions. In the present situation this would imply 
that on $dS_3\times S^1$ with antiperiodic boundary conditions for the
fermions, large $N$ correlators in the ${\mathbb Z}_N$
symmetric phase ($r_\x >
R_{\rm AdS}/2\sqrt 2  $) 
should be independent of the radius of the $S^1$.  In particular
then, for perturbations which are homogeneous along the circle, the
correlation functions should be independent of $r_\x$ and should match
up with the result on $dS_3\times {\mathbb R}$. The latter is obtained
by a (double) Wick rotation of $S^3\times {\mathbb R}$.
Since $\Tr F^2$ is a chiral primary in the ${\cal N}=4$ 
theory and its correlator on $S^3\times {\mathbb R}$ is not renormalized
by interactions, one would expect this to be true also on $dS_3\times
{\mathbb R}$.

\subsection{The massive case}
The holographic calculation of correlation functions in the
topological AdS black hole can be easily extended to massive
scalars. In the context of the Type IIB theory, such massive states are
stringy excitations with masses $M^2 \sim \alpha'^{-1}\gg R^{-2}_{\rm
  AdS}$. 
A scalar field of mass $M$ in the bulk is dual to a scalar operator
${\cal O}_\Delta$ in the field theory with conformal dimension
$\Delta=2+\sqrt{4+(MR_{\rm AdS})^2}$. We will study below the free massive
scalar in the bulk geometry to extract information on the analytic
structure of correlators of high dimension operators in the field
theory. 

There are two primary motivations for looking at high dimension operators
in the field theory: i) The works of
\cite{Fidkowski:2003nf,Festuccia:2005pi} have demonstrated that
propagators of heavy fields, in the geodesic approximation, may be
used to probe the bulk geometry behind horizons and perhaps 
extract information on singularities behind such horizons. 
ii) One of our main goals is to look for signatures of the
transition between a ${\mathbb Z}_N$ symmetric phase and a ${\mathbb
  Z}_N$ broken phase. In the bulk theory, the latter phase is the
small bubble-of-nothing geometry. Correlators in this latter geometry
can only be computed using an eikonal (WKB) approximation involving high
frequencies and/or large masses.

Extending the holographic analysis done above for massless fields, to
massive scalars in the topological black hole geometry,
we find that the  frequency space correlator is
\EQ
{
\tilde G_R(\nu)=  {\cal C}_\Delta\;
{\Gamma\left(\tfrac{1}{2}\left(\Delta-1-i\nu R \right)\right)^2 
\Gamma\left(3-\Delta\right)\over
\Gamma\left(\tfrac{1}{2}\left(3-\Delta-i\nu R\right)\right)^2
\Gamma(\Delta-1)},
\label{massive}
}
where the normalization constant ${\cal C}_\Delta = 2(\Delta -2)
\epsilon^{2(\Delta-4)}$, with $\epsilon \rightarrow 0$ as the boundary
is approached.
\begin{figure}[h]
\begin{center}
\epsfig{file=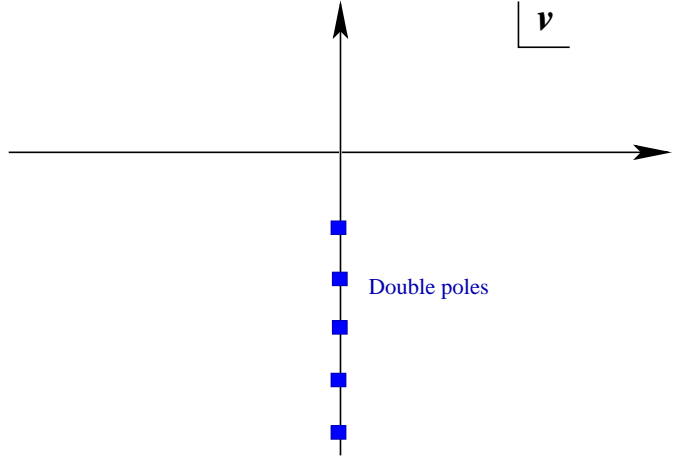,width=3.5in} 
\end{center}
\caption{\footnotesize Double poles in the lower half plane
for the massive (retarded) propagator at zero spatial momentum.}
\label{masstbh}
\end{figure}

In the massless limit $MR\rightarrow 0$, this reproduces the
expression \eqref{homog} found previously, 
after subtracting an additional divergent contact
term. The massive correlator has an analytic structure that is
qualitatively distinct from the massless case. In particular the
retarded correlator has an infinite set of {\em double poles} and
{\em simple poles} at
\EQ
{
\nu_k = -i (\Delta-1+2k){1\over R}\,;\qquad k\in {\mathbb Z}.
}
The significance of the appearance of double poles in the massive
retarded propagator is not entirely clear. Such double poles have also 
appeared in 2d CFT correlators with non-integer conformal dimensions from
the BTZ black hole \cite{Son:2002sd}, at zero spatial momentum. At
finite spatial momentum (e.g. $n\neq 0$) we expect the double
poles to split into simple poles. 

For real frequencies, the massive correlator also has an imaginary
part
\EQ
{
{\rm Im}\;\tilde G_R (\nu)= - \tfrac{\pi^2}{2} {\cal C}_\Delta 
{\Gamma(3-\Delta)\over \Gamma(\Delta-1)}\;
{\sin\left(\pi\,\Delta\right)\;\sinh\left(\pi \;\nu R \right)\over
|\Gamma\left(\tfrac{1}{2}(3-\Delta-i\nu R)\right)
\cos\left(\tfrac{\pi}{2}(\Delta-i\nu R)\right)|^4}.
\label{immassive}
}
The (de Sitter) thermal origin of this result is not as explicit as
for the massless scalar. However, after identifying the de Sitter
Hawking temperature to be $T_H=R^{-1}/2\pi$, it is worth comparing
the above expression with the imaginary part of the propagator in two
dimensions for
large non-integer conformal dimension deduced from  
the non-extremal BTZ black hole \cite{Son:2002sd}. The similarities
between the two results, particularly the numerator of 
\eqref{immassive}, are striking.

The large mass, high
frequency limit of this result is easily obtained, using Stirling's
approximation  
\EQ
{
\Gamma(z)\big|_{z\gg 1}\simeq \sqrt{2\pi} {1\over \sqrt z}\;e^{-z} \; z^z .
 }
When the masses are taken to be large so that $MR \gg 1$, then
$\Delta \approx M R$. In this high
frequency, large mass limit it is useful to define a rescaled
frequency variable 
\EQ
{
\tilde\nu \equiv {\nu \over M};\qquad \nu R,\; MR\gg 1,
}
so that
\EQ
{
G_R(\tilde\nu)\sim {\cal C}_\Delta \left({1- i
  \tilde\nu\over 2}\right)^{MR (1-i \tilde\nu)}\;\left({-1-i\tilde\nu\over 2}\right)^{MR(1+i\tilde\nu)}.
\label{largemass}
}
Here we have ignored an overall (real) phase due to frequency
independent coefficients in the large mass limit. At first sight, a
potentially problematic feature of this 
approximation is that there is a branch point singularity at $\tilde\nu=+i$ which
would imply a non-analyticity in the upper half plane, inconsistent
with the definition of a retarded propagator. Note that this 
feature is purely a result of the high frequency approximation
and the exact result \eqref{massive} has no singularities in the upper
half of the complex frequency plane. 
Indeed, closer
inspection reveals that the putative branch cut originating 
at $\tilde\nu = + i$ has a vanishing
discontinuity in the limit of large $MR$. The spurious branch cut
originates from the equally spaced zeroes of $G_R(\nu
)$ appearing
to coalesce in the high frequency limit. The branch cut
discontinuity at $\tilde\nu=-i$ is, however a genuine non-analyticity and
originates from the infinite set of poles merging into a continuum in
the high frequency approximation.
\begin{figure}[h]
\begin{center}
\epsfig{file=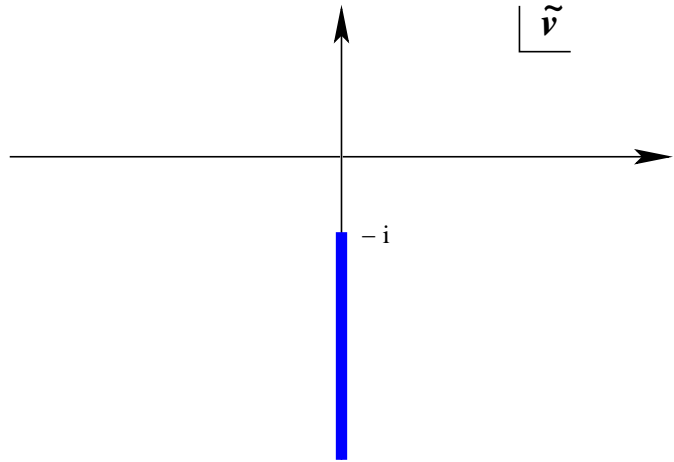,width=3.5in}
\end{center}
\caption{\footnotesize For large mass and frequency, the
propagator in the topological black hole phase, has a branch cut as
indicated in the rescaled frequency plane, $\tilde\nu=\nu/M$. This
results from the apparent merging 
of the infinite set of isolated poles of the exact
Green's function.} 
\label{largetbh}
\end{figure}
The easiest way to understand the singularities and discontinuities of
the function above, is to examine the function $z^ {2 MR z}
\;(z-1)^{-2 MR(z-1)}$ and then make the replacement 
$z \rightarrow (1-i\tilde\nu)/2$. 

Along the branch cut ${\cal B} \equiv \left\{\tilde\nu \in [-i,
  -i\infty)\right\}$, we find that the retarded propagator has a
discontinuity
\SP
{
{\rm Disc}|_{\cal B} \;\tilde G_R(M\tilde\nu)= &2 i \;{\cal C}_\Delta
\;\sin(2\pi MR)
\left({|\tilde\nu|+1\over 2}\right)^{MR(|\tilde\nu|+1)}
\left({|\tilde\nu|-1\over 2}\right)^{-MR(|\tilde\nu|-1)}
\\\nonumber
& \tilde\nu = -i |\tilde\nu|\,, \quad |\tilde\nu|\geq 1.
}
It is clear that the discontinuity is large in the large mass limit.

Now let us look closely at the putative branch cut along the imaginary
axis ${\cal B}' \equiv \left\{\tilde\nu \in [-i, i]\right\}$. Computing the
discontinuity across this, we have
\SP
{
{\rm Disc}|_{{\cal B}'} \;\tilde G_R(M\tilde\nu)&= 2 i\; {\cal C}_\Delta
\sin\left(MR\pi(1- x)\right) \left({1+x\over 2}\right)^{MR(1+x)}
\left({1-x\over 2}\right)^{MR(1-x)};\\
&\tilde\nu = i x,\,\quad -1\leq x\leq 1.
}
This vanishes when $MR \rightarrow \infty$, for two reasons. 
The rapid sinusoidal oscillations will give vanishing contribution
to any contour integral along ${\cal B}'$. Furthermore the amplitude
of the oscillations vanishes exponentially as evident from the
expression above.

In the leading high frequency approximation , for real frequencies, the
imaginary and real parts of the Green's function are given by
\SP
{
\tilde G_R(\nu)
&\approx {\cal C}_\Delta\;\left({\nu^2+M^2\over 4M^2}\right)^{MR}
e^{-\tfrac{1}{\pi} \tfrac{\nu}{T_H}\tan^{-1}\tfrac{\nu}{M}}\; 
e^{-\tfrac{|\nu|}{2T_H}}\left(\cos(\tfrac{M}{2T_H})- i\; 
{\rm sgn}(\nu)\sin(\tfrac{M}{2T_H})\right),\\
&T_H={R^{-1}\over 2\pi}\,,\quad\nu\in{\mathbb R}.
}
The high frequency, large mass limit thus appears to retain features
of the thermal effects of de Sitter space.  
This result can be deduced from \eqref{largemass} after choosing an
appropriate branch of the function and also directly follows from 
\eqref{immassive}. We will see subsequently that these high frequency
expressions can also be derived by solving the wave equations using 
a WKB approximation, providing a consistency check.

\subsection{R-current correlation functions}
Real time response to perturbations of a conserved charge density can
reveal interesting late time physics, such as hydrodynamic or
diffusive relaxation. It is now well understood and 
established \cite{Policastro:2002se, Son:2007vk} via holographic
calculations in AdS 
black hole backgrounds, that correlators of conserved global currents
exhibit hydrodynamic and diffusion poles. R-charge diffusion in the
strongly coupled, high temperature ${\cal N}=4$ plasma was first
discovered in \cite{Policastro:2002se}. The universal features of
strongly coupled plasmas follow from the properties of the stretched
horizon of AdS black holes \cite{Son:2007vk}. 
It is therefore natural to ask whether the
horizon in the topological AdS black hole geometry implies 
hydrodynamic behaviour of correlation functions in the dual field
theory. The answer to this question will depend on the relevant time scales
involved since the boundary field theory is formulated on an expanding
background, namely $dS_3\times S^1$.

The strong coupling correlators for $SO(6)$ R-currents of ${\cal N} = 4$ SUSY
Yang-Mills theory on $dS_3 \times S^1$ will be obtained holographically from
the on-shell action for the $SO(6)_R$ gauge fields in the topological
$AdS_5$ black hole background, by
following the prescription of \cite{Policastro:2002se}. The Maxwell
action for the gauge field is 
\begin{equation}
S = -\frac{1}{4 g^2_\mathrm{SG}} \int d^5 x \sqrt{-g} g^{\m \a} g^{\n
  \b} F_{\m \n} F_{\a \b}, \label{action} 
\end{equation}
varying which gives the following equations of motion
\begin{equation}
\frac{1}{\sqrt{-g}} \partial_\n \left( \sqrt{-g} g^{\m \a} g^{\n \b}
  F_{\a \b} \right) =0 \label{eom}. 
\end{equation}
Here, we have $g_\mathrm{SG}^2 = 16 \p^2 R_{\rm AdS} \slash N^2$. We
find it convenient to use the following form for the metric in the
region exterior to the horizon,
\begin{equation}
ds^2 = \frac{R^2_{\rm AdS}}{4 z^2 \left(1-z\right)} dz^2 + \frac{R^2_{\rm AdS}
  (1-z)}{z}\left[-d \t^2 + \cosh^2 \t \;d \W_2^2 \right]+\frac{r_\x^2}{z} \;d
\x^2.
\end{equation}
Substituting
the solution to the equation of motion that corresponds to
the boundary value $A_{\a}(z)\big|_{z=0} = A_{\a}^0$ back into the
gauge field action, we will get a generating functional for the
R-charge correlators of the field theory.   

Since the field theory lives on the $dS_3\times S^1$
boundary with spatial $S^2\times S^1$ spatial slices,  
it is natural to consider the late time behaviour of
long-wavelength fluctuations in the following two cases:
\begin{enumerate}
\item The R-charge perturbation is inhomogeneous on the $S^1$, but
  homogeneous on the spatial section of $dS_3$, 
\item The fluctuation is homogeneous on the circle, but inhomogeneous
  on the $S^2$. 
\end{enumerate}
Of particular interest is the presence or the lack of late time hydrodynamic
relaxation of this system.

\subsubsection{Inhomogeneous perturbation on the $S^1$ \label{S1}}
In the first case, we will assume, for simplicity 
that the R-charge perturbation carries no momentum around the 
$S^2$. Furthermore, we use gauge freedom to set the radial component
of the gauge potential $A_z = 0$. Hence, from the symmetries of the
configuration, there are two
remaining bulk gauge fields that are non zero:
\EQ
{
A_\tau=A_\tau(z,\tau,\x), \qquad A_\x = A_\x(z,\tau, \x).
}
Since the $\x$-direction is a spatial circle, the two components of
the vector potential,
$A_\tau$ and $A_\x$, can be conveniently expanded in Fourier modes 
on the circle. The time dependence can also be re-expressed in terms
of a mode expansion. There is a subtlety involved in this however.
For $A_\x$, which is a scalar in $dS_3$, the mode decomposition 
is straightforward
\EQ
{
A_\x(z,\t,\x)= \sum _n \frac{e^{i n \x}}{2\pi} 
\int_{-\infty}^\infty 
\frac{d\nu}{2\pi} \,\,{{\cal T}_\x (\nu,\t)}\;\;{\cal G}_n(\nu,z),
}
where $n \in \mathbb{Z}$. On the other hand $A_\t$, which is a gauge field
in $dS_3$, has its complete time dependence captured by 
normal modes of two kinds -- normalizable and delta-function
normalizable. In fact, below we show that 
there is a single normalizable mode and a
continuum of delta-function normalizable states obtained as solutions
to a Schr\"odinger problem. Anticipating this we write
\EQ
{
 A_\tau(z,\t,\x)= \sum _n \frac{e^{i n \x}}{2\pi} 
\left(\int_{-\infty}^\infty \frac{d\nu}{2\pi} \,{\cal T}_\tau (\nu,\t)\;  
{\cal F}_n(\nu,z) + {\cal T}_\t^{\rm N}(\t) \,{\cal F}^{{\rm N}}_n(z)\right),
}
where ${\cal T}_\t(\nu,\t)$ and ${\cal T}^{\rm N}_\t(\t)$ will be the
delta-normalizable and normalizable modes respectively.
The mode functions ${\cal T}_{\x,\t}$ are solutions to 
\AL
{&\dd{}{\t}\left({\cosh^{-2} \t} \,\dd{}{\t} \left(\cosh^2 \t \, {\cal T}_\t
    (\nu,\t) \right) \right)
= - (\nu^2 +1) \;{\cal T}_\t (\nu,\t), \\\nonumber
&{\cosh^{-2}\t}{d\over d\t}\left(\cosh^2\tau\,{\cal T}_\tau\right)= -i
(\nu+i){\cal T}_\chi.
}
The first of these can be put in the form of a Schr\"odinger equation,
by defining ${\cal T}_\t = \tilde{\cal T}/\cosh\t$, 
\EQ
{
-\frac{d^2}{d\tau^2}\tilde {\cal T} -{2\over \cosh^2\tau}\tilde{\cal T}
=\nu^2\,\tilde {\cal T}.
}
It is solved by the associated Legendre
function ${P}_1^{-i\nu}(\tanh \tau)$. 
For $\nu^2\geq 0$, there is a
continuous infinity of delta-function normalizable states, which yield
\EQ
{
{\cal T}_\t (\nu,\t) = 
%\sqrt{2 \pi} 
\, \frac{e^{- i \nu \tau} }
{\cosh \t}\left(\frac{\nu-i\tanh\t}{\nu-i}\right), \qquad\quad
{\cal T}_\x(\nu,\t) =  \frac{e^{- i \nu \tau} }{\cosh \t},\qquad \nu^2
\geq 0.
\label{modes}
}
The Schr\"odinger potential above also has bound states for $\nu^2 <
0$. In fact there is precisely one normalizable bound state with $\nu^2=-1$,
corresponding to the solution
\EQ
{
{\cal T}_\t^{\rm N}(\tau)= {1\over\sqrt 2}\,{1\over \cosh^2\tau}.
}
This solution can be directly obtained by evaluating $P_1^{-i\nu}(\tanh\t)$ at
$\nu^2= -1$ (and appropriately normalized) or can be systematically inferred 
from the Maxwell equations.
The continuum modes for $A_\x$ and $A_\t$
are orthonormal with respect to the inner product 
\begin{equation}
\langle {\cal T}_{a},{\cal T}_{a} \rangle  \equiv
\int_{-\infty}^\infty d\t \cosh^2 \t \; {\cal T}_{a}(\nu,\t) {\cal
  T}_{a}(\nu',\t) = 2 \pi \,  \d(\nu + \nu'), 
\end{equation}
for $a \in \{ \x,\t \}$, while the bound state is normalized so that
\EQ
{
\int_{-\infty}^\infty d\t \,\cosh^2\t\,\left({\cal T}^{\rm N}_\t\right)^2=1.
}

In the far future the continuum modes are both given by ${\cal T}_a \sim
e^{-i\nu\tau} e^{-2\tau}$. Upon analytically continuing to 
the complex $\nu$ plane, for $\nu = i +\w$, with $\w \in {\mathbb R}$, they
are propagating (purely oscillatory) excitations with frequency $\w$ in the far
future. On the other hand the (normalizable) bound state decays
exponentially in the far past and future and being real does not
contribute to the flux at the horizon of the topological black hole.

Using these modes to eliminate the $\t$-dependence we find
three equations that depend only on ${\cal F}_n(\nu,z)$ and ${\cal
  G}_n(\nu,z)$  
\AL
{&-\tfrac{1}{R^2_{\rm AdS}}\,(\nu+i) \;{\cal F}_n'+ \frac{n (1-z)}{r_\x^2}
  \,{\cal G}_n'= 0,   
\label{cons}\\
&\tfrac{1}{R^2_{\rm AdS}}\dd{}{z}\left((1-z) {\cal F}_n' \right) + 
\frac{n}{r_\x^2 
}\,(\nu-i) \,{\cal G}_n - \frac{n^2}{r_\x^2} \,{\cal F}_n =0, \label{eq3}\\  
&4 z\dd{}{z}\left((1-z)^2 {\cal G}_n' \right) - n(\nu+i) {\cal F}_n
+(\nu^2+1) {\cal G}_n= 0.
\label{eq4} 
}
Here, prime denotes a derivative with respect to $z$. The radial or
$z$-dependence of the bound state solution, ${\cal F}^{\rm N}(z)$ 
is found by analytically
continuing the profile for generic $\nu$ to $\nu=\pm i$.
Note that there
are three equations for two unknowns; so, to ensure a non-trivial
solution any two equations must imply the third and it is
straightforward to check that this is indeed the case. We  can then
use these equations of motion to derive two independent ones, 
each containing only one of the  unknown functions:
\AL
{
&4z (1-z) {\cal F}_n''' -4 (3z-1) {\cal F}_n''-4 {\cal F}_n' = \left[
  \bar{n}^2-\frac{\nu^2+1}{1-z} \right] {\cal F}_n',
\label{gaugea}\\
&4 z (1-z) {\cal G}_n''' - 4(5z-1) {\cal G}_n''-
\frac{8(1-2z)}{1-z}{\cal G}_n' = \left[\bar{n}^2 -
  \frac{(\nu^2+1)}{1-z}\right]{\cal G}_n', \label{gaugeb}
}  
where we have defined $\bar{n} \equiv \frac{n R_{\rm AdS}}{r_\x}$. These
equations are immediately solved in terms of hypergeometric
functions. Singling out the 
solutions that satisfy the purely infalling wave boundary condition at
the horizon, we find the induced boundary action for the Maxwell fields 
(see Appendix A for details). 
The R-current correlators can be read off
from the finite and non-analytic pieces of this boundary action
\eqref{bulkmaxwell1}, \eqref{bulkmaxwell2}. We will denote the Fourier harmonics of the
R-current along the spatial circle as
\EQ
{
j_n^\mu(\tau) = \int_0^{2\pi}d\x \, e^{-i n\x}\, j^\mu(\x,\t),
}
where we have already restricted attention to the s-wave sector on the
spatial two-sphere. For perturbations with non-vanishing momentum
along the spatial circle, we define the retarded, real time, Green's
functions as  
\EQ
{G^{\mu\nu}(\t,\t'; n)=\langle[\,j_n^\mu(\tau),\,j_{-n}^\nu(\tau')]\rangle\,
\Theta(\tau-\tau')\,,\qquad \mu,\nu \in \{\x,\t\}.
}
The conserved global currents are in one-to-one correspondence with
the boundary values of the 5-d gauge fields. Functionally
differentiating the induced boundary action \eqref{bulkmaxwell2} 
with respect to the boundary values of the gauge fields, we find
\AL
{G^{\t \t}(\t,\t';n) = &\qquad\,
\cosh^2\t\cosh^2\t'\times \label{gtt}
\\\nonumber
&n^2\,\left[\int_{-\infty}^\infty {d\nu\over 2\pi}\,{\cal T}_\tau(\nu,\tau){\cal
  T}_\tau(-\nu,\t') \,\,{\bf \Xi}(\nu,n) 
+ {\cal T}_\t^{\rm N}(\t){\cal T}_\t^{\rm
  N}(\t')\,\,{\bf \Xi}(i, n)
%\times
%\right.\\\nonumber
%&\left.
%\left(\psi\left(1+\tfrac{i}{2}\bar n\right) +
%\psi\left(1-\tfrac{i}{2}\bar n\right)\right)
\right], \\\nonumber\\
G^{\x \x}(\t,\t';n) =&
\,\cosh^2\t\cosh^2\t'
\int_{-\infty}^\infty {d\nu\over 2\pi}\,{\cal T}_\x(\nu,\t)
{\cal T}_\x(-\nu,\t')\left(\nu^2 + 1\right)\,{\bf \Xi}(\nu,\,n), \\\nonumber\\
G^{\t \x}(\t,\t';n) = &\,G^{\x \t\,*}(\t,\t';n) 
= &\\\nonumber
& \cosh^2\t\cosh^2\t'\,n\int_{-\infty}^\infty{d\nu\over 2\pi}\,{\cal T}_\t(\nu,\t)
{\cal T}_\x(-\nu,\t')\left(\nu -i\right)
{\bf \Xi}(\nu, n),
\\\nonumber\\
{\bf \Xi}(\nu,\,n)= &\tfrac{N^2 R_{\rm AdS}}{32 \pi^2 r_\x}\left(\psi\left(\tfrac{1}{2}+\tfrac{i}{2}(\bar  
  n-\nu)\right) +
\psi\left(\tfrac{1}{2}-\tfrac{i}{2}(\bar
  n+\nu)\right)\right).
}

It is easily established that the above expressions are indeed
retarded Green's functions and are non-vanishing only when $\t >
\tau'$. Two essential features ensure that
this is the case: The function ${\bf \Xi}(\nu, n)$ appearing
universally in the $\nu$-integrals has only simple poles in the lower half
complex plane at
\EQ
{
\nu= -i(2k +1) \pm \bar n\,,\qquad k\in {\mathbb Z}.
}
There is a second source of non-analyticities in the $\nu$-plane. This
lies in the $\nu$-dependent normalization of the mode functions ${\cal
T}_\t(\nu,\t)$ \eqref{modes}. The potentially worrisome aspect of this
is the appearance of a pole in the {\em upper} half plane at $\nu=+i$, which
gives a non-vanishing contribution for $\t < \t'$. However, the
potential problem is eliminated by the term dependent on the discrete,
normalizable mode
${\cal T}^N_\t$ in \eqref{gtt} which exactly cancels against the
contributions from the poles at $\nu =+i$, ensuring that our
Green's function is causal. The remaining Green's functions are
manifestly free of singularities in the upper half plane.

We thus see that the Son-Starinets recipe for determining real time
response functions works in the case of the topological black hole, 
provided we carefully account for the
contributions from both the continuum and discrete mode functions in $dS_3$.

It is also clear in the above expressions, that there are no diffusion
poles. Instead, from the properties of the digamma function which we
have encountered before, the frequency space Green's function which is
effectively ${\bf \Xi} (\nu,n)$\,, has only simple poles in the complex
$\nu$-plane, the lowest of these being at 
\EQ
{
\nu= -i\pm \,\bar n.
} 
Excitations with complex $\nu = \w - i$, and $\w\in {\mathbb R}$ 
will propagate at late times, the perturbations simply
evolving as left- and right-moving excitations on the
$S^1$ without dissipating. 

The absence of any diffusion or transport like behaviour may be
intuitively explained by noticing the similarity of the topological
AdS black hole to the BTZ black hole %\footnote{As a matter of fact,
  %references \cite{Banados:1997df, Banados:1998dc} studied the topological
  %black hole as the higher-dimensional analogue of the BTZ black hole.}. 
  Setting up an excitation with momentum only on the $S^1$ is
equivalent to saying the variation of the fields along the $S^2$ is
zero. Therefore, aside from the time dependent factors associated to the de
Sitter expansion, the metric that is ``seen'' by the bulk fields is not
the full five dimensional metric, but its effective $2+1$
dimensional portion, 
\begin{equation}
ds^2 = -R^2_{\rm AdS}
(\rho^2-1) d \t^2 +\frac{R^2_{\rm AdS}}{\rho^2-1} d\rho^2  + \rho^2 r_\x^2 d \x^2
\end{equation}
Comparing this with the metric for a $2+1$ dimensional BTZ black hole
with zero angular momentum 
\begin{equation}
ds^2 = -\left(\frac{r^2}{R^2} - M \right) dt^2 + \left(\frac{r^2}{R^2}
  - M \right)^{-1} dr^2 + r^2 d \phi^2, 
\end{equation}
we note the obvious similarity. Therefore we expect the behaviour of
fields in the 
topological black hole background with an inhomogeneous excitation
around the $S^1$, to be similar to the behaviour of the fields in a BTZ
black hole background.  In other words, they should behave as in a
(1+1)-dimensional CFT \cite{Son:2002sd}, just as we see from our
results above. 

\subsubsection{Inhomogeneous perturbation on the $S^2$ \label{dS3}}

We will now examine the real time response to fluctuations carrying
momentum along the spatial sections of three dimensional de Sitter space.
Each spatial section of $dS_3$ is a two-sphere which undergoes
exponential expansion at late times. For this case, we will 
focus on a situation where an inhomogeneous R-charge perturbation is
set up on the two-sphere 
with only a dependence on the polar angle $\theta$ and time
$\tau$. For this configuration, the dual bulk gauge fields are
\EQ
{
A_\tau=A_\tau(z,\tau, \o)\,;\qquad A_\theta = A_\theta(z,\tau, \o)\,;\qquad 
A_z=A_\chi=0,
}
where $A_z$ is set to zero by the gauge freedom, and $A_\chi$
vanishes due to $\chi$-independence of the configuration. By spherical
symmetry, the scalar potential $A_\tau$ and the vector 
potential $A_\theta$, each can be expanded in terms of scalar
and vector spherical harmonics, respectively:
\EQ
{
A_\tau = \sum _{\ell=0}^\infty
{\cal F}_\ell(z,\tau)\;Y_\ell^0(\theta), \qquad
A_\theta=\sum_{\ell=1}^\infty {\cal G}_\ell(z,\tau)\;\partial_\theta
Y_\ell^0(\theta). 
}
Substituting these into the bulk Maxwell equations, we find
\AL
{
&4 z(1-z){\partial_z}\left ((1-z){\partial_z
  {\cal G}_\ell}\right) -{\partial^2_\tau {\cal G}_\ell}
+{\partial_\tau {\cal F}_\ell}=0
\label{atheta}
\\\nonumber\\
& 4 z(1-z)\partial_z \left ((1-z){\partial_z
  {\cal F}_\ell}\right) - {\ell(\ell+1)\over \cosh^2 \tau}
\left({\cal F}_\ell -\partial_\tau {\cal G}_\ell\right)=0,
\label{atau}
\\\nonumber\\
&\partial_\tau \left(\cosh^2\tau\partial_z {\cal F}_\ell\right) +
\ell(\ell+1)\partial_z {\cal G}_\ell=0.
\label{conserve}
}
>From equations \eqref{atau} and \eqref{conserve} we obtain a 
differential equation for ${\cal F}'_\ell \equiv \partial_z {\cal F}_\ell$
\EQ
{
4 \partial_z \left(z(1-z)\partial_z\left((1-z){\cal F}'_\ell\right)\right)
-{\ell(\ell+1)\over \cosh^2\tau} {\cal F}'_\ell
-{1\over\cosh^2\tau}\partial_\tau^2 \left(\cosh^2\tau {\cal F}'_\ell\right)=0.
\label{radial}
}
Now we will separate out the explicit temporal dependence, keeping in
mind, as before, the possibility of contributions from 
both discrete and continuous modes
\EQ
{
{\cal F}'_\ell(z,\tau)=\int_{-\infty}^\infty{d\nu}
\; {\cal T}_\ell(\nu,\t) \;F_\ell(\nu,z) + 
\sum_{m}{\cal T}_{\ell\,m}^{\rm N}(\t)\, F_{\ell \,m}^{\rm N}(z).
}
The mode functions ${\cal T}_\ell$ satisfy
\EQ
{
\left[- \partial_\tau^2 -
{\ell(\ell+1)\over \cosh^2\tau} \right]
\left(\cosh^2\tau\,{\cal T}_\ell\right) = \nu^2 \;
\left(\cosh^2\tau\,{\cal T}_\ell\right) ,
\label{pt}
}
which is a Schr\"odinger equation whose potential clearly will have
both bound states and scattering or continuum states. The full set of
solutions form an orthonormal, complete set. Indeed, the
delta-normalized eigenstates are the Legendre functions 
\EQ
{
{\cal T}_\ell(\nu,\t)= 
\Gamma(1+i \nu)\,{P_\ell^{-i \nu}(\tanh\tau)\over \cosh^2\tau}.
}
Those with $\nu^2>0$ are scattering states with continuous values of 
$\nu \in {\mathbb R}$, while the
discrete, ``bound states'' have $-i \nu =1,2,\ldots \ell$,
\EQ
{
{\cal T}_{\ell\,m}^{\rm N} 
= \sqrt{m(\ell-m)!\over (\ell+m)!}
\,\,{P_\ell^m(\tanh\tau)\over \cosh^2\tau}\,,\qquad m=1,2,\ldots \ell.
}

For $\nu^2> 0$, the late time, $\tau \to\infty$, behaviour
of the modes will be significant, 
\EQ
{
{\cal T}_\ell(\nu,\t)\big|_{\tau\gg 1} \to e^{-i \nu\tau}\; e^{-2\tau},
}
as these modes are oscillatory.  Applying infalling boundary
conditions on these modes at the
horizon of the topological AdS black hole, \eqref{radial} yields,
\EQ
{
F_\ell(\nu,z)= C_\ell(\nu)\; (1-z)^{-1-i\nu/2}
\;{}_2 F_1\;\left(-i\tfrac{\nu}{2},1-i\tfrac{\nu}{2}\,;1-i\nu\,;1-z\right).
\label{gaugesol}
}
For the discrete series, the radial profile in the bulk, 
$F_{\ell \,m}^{\rm N}(z)$ is obtained by evaluating $F_\ell(\nu,z)$ at
$\nu= -i m$. 
Putting the above ingredients together, 
the general form of the electric field along the
radial direction in the bulk is 
\SP
{
A_\tau'(z,\tau,\o)=&
\sum_{\ell=0}^\infty Y_\ell^0(\theta)\left(\int 
_{-\infty}^\infty\;{d\nu\over 2\pi }\, {\Gamma(1+ i \nu)}\,
{P_\ell^{-i \nu}(\tanh\tau)\over \cosh^2\tau}\;
F_\ell(\nu,z)+\right.\\
&\left.+\sum_{m=1}^\ell
\sqrt{m\tfrac{(\ell-m)!}{(\ell+m)!}} \, {P_\ell^m(\tanh\t)\over
  \cosh^2\t}\, F_{\ell \,m}^{\rm N}(z)
\right).
}
This also allows us to automatically solve for $A_\theta'$
using \eqref{conserve} and we get
\SP
{
A_\theta'(z,\tau,\theta)= &- \sum_{\ell=1}^\infty
{\partial_\theta Y_\ell^0(\theta)\over
  \ell(\ell+1)}\,\left(\int_{-\infty}^\infty {d\nu\over 2\pi}\; 
\Gamma(1+i\nu)\,\partial_\tau P_\ell^{-i \nu}(\tanh\tau)\;
F_\ell(\nu,z)+\right.\\
&\left.+ \sum_{m=1}^\ell
  \sqrt{m\tfrac{(\ell-m)!}{(\ell+m)!}}
\,\,\partial_\t P_\ell^m(\tanh\t)\,F_{\ell\, m}^{\rm N}(z)
\right).
}
Now, the bulk gauge field action can be shown to induce a boundary term which
will be the generating functional for the boundary R-current
correlators. Using the explicit solutions above, the induced boundary
action becomes
\EQ
{
S=  {1\over 2 g^2_{SG}}\int d\tau \;
r_\x\left[\sum _{\ell=0}^\infty 
{\cal F}_\ell(z,\tau)\, {\cal F}'_\ell(z,\tau) 
\cosh^2\tau + \sum_{\ell=1}^\infty 
\ell(\ell+1)\,{\cal G}_\ell(z,\tau)\, {\cal G}'_\ell(z,\tau)
\right]_{z=\epsilon\to 0}.
\label{bdryaction}
}
The next step is to express this completely in terms of the boundary values of
the gauge potentials 
${\cal F}_\ell^0(\t) \equiv {\cal F}_\ell(\epsilon,\tau)$ and ${\cal
  G}_\ell^0(\tau) \equiv {\cal G}_\ell(\e,\t)$.  Their radial derivatives 
${\cal F_\ell}'$ and ${\cal G}_\ell'$
(equivalently $A_\tau'$ and $A_\theta'$ ) at the boundary
$z=\epsilon$, are also determined completely by the 
boundary values of the gauge potentials, ${\cal F}_\ell^0(\t)$ and
${\cal G}_\ell^0(\tau)$ as in \eqref{derivatives}.

>From the boundary action above, we thus find that the real time, 
retarded Green's functions for the R-charge currents $j^{\mu}$, in the
gauge theory are 
\AL
{
G^{\tau\tau}(\tau,\t'; \ell)=&  
\,\ell(\ell+1)
\left[\int_{-\infty}^\infty  
{d\nu\over 2\pi} \; \,{\pi \nu\over \sinh\pi\nu}\; 
P_\ell^{-i\nu}(\tanh\tau)P_\ell^{i\nu}(\tanh\tau')\,{\bf \Upsilon}(\nu)+
%\left(\psi(-\tfrac{i\nu}{2}) - \tfrac{1}{i \nu}\right),
\right.\\\nonumber
&\left.\qquad\quad+ \sum_{m=1}^\ell \, (-1)^m\,m \,
P_\ell^m(\tanh\t)P_\ell^{-m}(\tanh\t')\,{\bf
    \Upsilon}(i\,m)\right], \\\nonumber\\
G^{\theta\theta}(\t,\t';\ell)=&\,\ell(\ell+1) 
\left[\int_{-\infty}^\infty  {d\nu\over 2\pi}\;{\pi \nu\over \sinh\pi\nu}\; 
\partial_\t P_\ell^{-i\nu}(\tanh\tau)\, \partial_{\t'}P_\ell^{
  i\nu}(\tanh\tau') {\bf \Upsilon}(\nu)+ \right.\\\nonumber
&\left.\qquad\quad+ \sum_{m=1}^\ell \, (-1)^m\,m\,
\partial_\t P_\ell^m(\tanh\t)\partial_{\t'}P_\ell^{-m}(\tanh\t')\,
{\bf
    \Upsilon}(i\,m)\right], \\\nonumber\\
G^{\tau\theta}(\tau,\t';\ell)=& \ell(\ell+1)\left[\int_{-\infty}^\infty 
{d\nu\over 2\pi}\;{\pi \nu\over \sinh\pi\nu}\; 
P_\ell^{-i\nu}(\tanh\tau)\, \partial_{\t'}P_\ell^{i\nu}(\tanh\tau')
\,{\bf \Upsilon}(\nu)+\right.\\\nonumber
&\left.  \qquad\quad+ \sum_{m=1}^\ell (-1)^m\, m\,
P_\ell^m(\tanh\t)\partial_{\t'}P_\ell^{-m}(\tanh\t')\,
{\bf \Upsilon}(i\,m)\right],\\\nonumber\\
&{\bf \Upsilon}(\nu)=\tfrac{N^2 r_\x}{64\pi^2 R_{\rm AdS}}\,
\left(\psi(-\tfrac{i\nu}{2}) - \tfrac{1}{i \nu})\right).
%\langle j_\o^\ell(\tau)j_\t^\ell(\tau')\rangle_R &= \frac{N^2
%  r_\x\,\ell(\ell+1)}{64\pi^2 R} \int {d\nu\;\pi \nu\over \sinh\pi\nu}\; 
%\partial_{\t}P_\ell^{-i\nu}(\tanh\tau)\, P_\ell^{i\nu}(\tanh\tau')
%\left(\psi(-\tfrac{i\nu}{2}) - \tfrac{1}{i \nu})\right). 
}
We need to first confirm that these Green functions satisfy basic
consistency checks. Specifically, the retarded functions must vanish
for $\t < \t'$.  As in the previous case, this property is not
manifest, but follows from the nature of the non-analyticities of
${\bf \Upsilon}(\nu)$, and the normalized mode functions 
$\Gamma(1+i\nu)P_\ell^{-i\nu}(\tanh\tau)$, in the complex $\nu$-plane. 
For real values of $\nu$, the associated Legendre function is
\cite{elliptic, Abramowitz:1964} 
\EQ
{
\Gamma(1+i\nu)\,P_\ell^{-i \nu}(\tanh\t)= e^{-i\nu\t}\,
{}_2F_1(-\ell,\,\ell+1;\, 1+ i\nu,\, (1-\tanh\t)/2).
%\times {\rm Polynomial}\,\,
%{\rm in}\,\, \tanh\t.
}
For integer $\ell$, the hypergeometric function is a finite polynomial in 
$\tanh\t$ and a ratio of degree $\ell$ polynomials of $\nu$.
Therefore the exponential frequency dependence means that, for $\t-\t' <  0$,
the integrals over the frequency $\nu$ can be  
evaluated by closing the contour in the upper half plane. The function 
${\bf \Upsilon}(\nu)$ has no poles in the upper half complex plane. It
has simple poles at 
\EQ
{
\nu= 0, \,-2i, \,-4i \ldots
}
The normalized modes, $\Gamma(1+i \nu) P_\ell^{-i\nu}(\tanh\t)$ have
exactly $\ell$ simple poles in the upper half plane at
$\nu=i,\,2i\ldots \ell i$. The contributions from these are, however,
cancelled by the inclusion of the discrete modes in the retarded
Green's function above. Hence our correlators are zero for $\t < \t'$.

\subsubsection{Late time behaviour}

The real time response functions in general contain important
information on the long time relaxation of perturbations away from the
equilibrium or ground state. In thermal field theory on flat space,
the relaxation of such fluctuations of conserved charges proceeds
via hydrodynamic or diffusion modes. The response functions at strong
coupling then exhibit diffusion poles in frequency space, $G^{\t\t}
\propto (i\omega-  D k^2)^{-1}$, where $\omega$ is the frequency and
$k$, the soft spatial momentum. Due to the explicit time dependence of
the background metric, we cannot do a similar frequency space study of
the full Green's functions in de Sitter space. Instead, we could analyze
their behaviour as functions of time.

In de Sitter space, perturbations labelled by wave number
$\ell$, get red-shifted so that given sufficient time 
their physical wavelengths become super-horizon sized. This happens when 
\EQ
{
{ \ell \, e^{-\tau}\over R} \sim {1\over R}.
}
At late enough times, even very high harmonics on the sphere get
stretched  and eventually exit the horizon. To zoom in on the time
evolution of such modes, it is useful to think of $\ell e^{-\tau}$,
the physical wave number, as being fixed as $\t \to
\infty$. For example, in this late time approximation
we neglect terms like $\ell \,e^{-2\tau}$
in comparison to powers of $\ell \,e^{-\t}$.
This is practically equivalent to going to planar coordinates
for de Sitter space and the mode functions behave as 
\EQ
{
P_\ell^{-i \nu}(\tanh\t)\big|_{\ell e^{-\t}=\,{\rm fixed}} \longrightarrow 
\ell^{-i \nu}\,J_{i\nu}\left(2\ell e^{-\t}\right)
\label{bessel}
}
It is possible to derive this by replacing the potential
$\ell(\ell+1){\rm sech}^2\t$ in the mode
equation \eqref{pt},  with $4\,\ell^2\,e^{-2\tau}$.
Note that, instead of a fixed physical wavelength if we focus
attention on fixed comoving wavenumber, given by $\ell$, all modes
simply approach the $s$-wave at late times,
\EQ
{
\lim_{\t \to \infty}\,\Gamma(1+i\nu)\, 
P_\ell^{-i\nu}(\tanh\tau)\big|_{\ell\,\,{\rm fixed}} 
\longrightarrow 
e^{-i\nu \t}.
}

\begin{figure}[h]
\renewcommand{\figurename}{Fig.}
\begin{center}
\epsfig{file=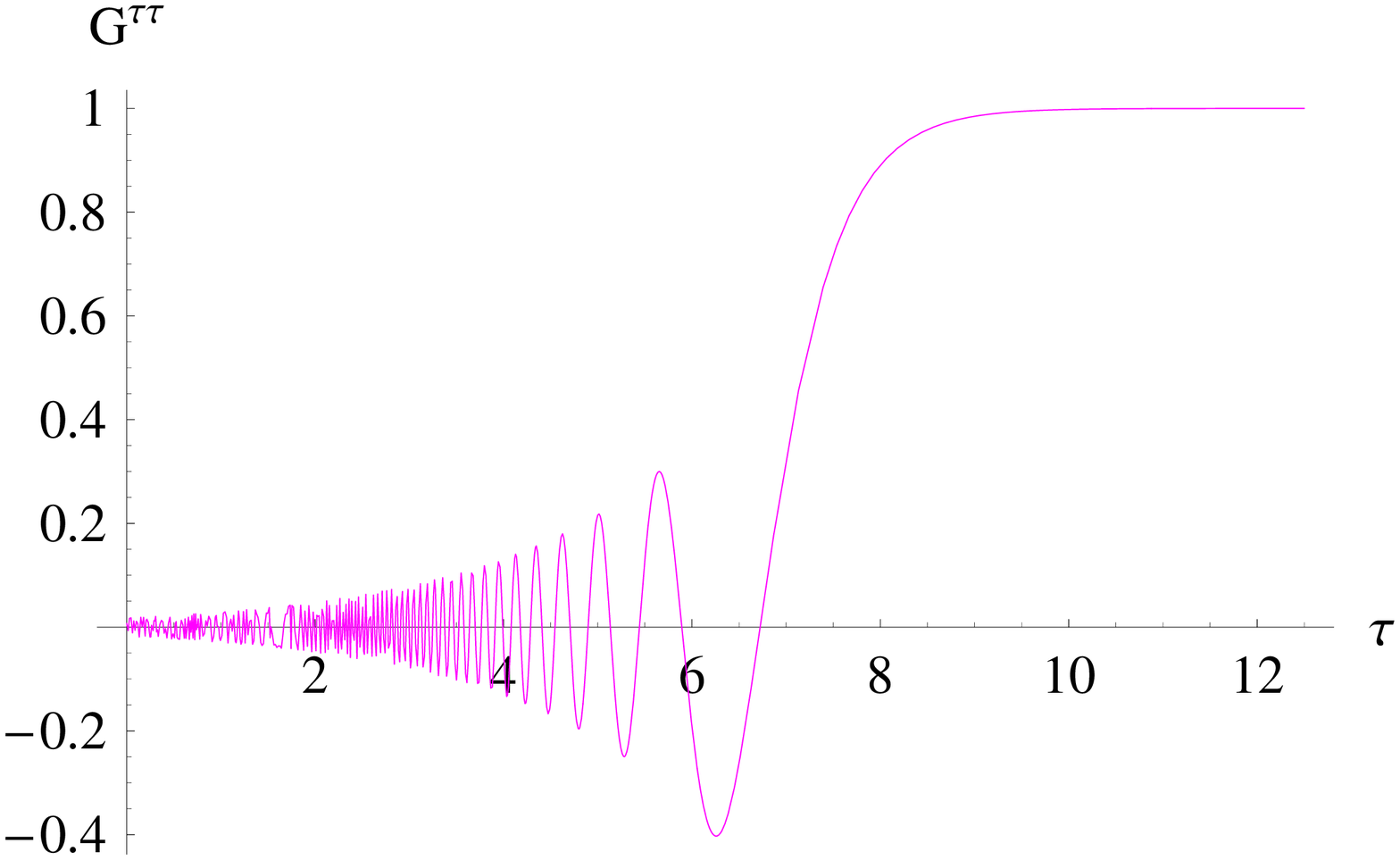,width=2.5in} \hspace{0.3in}
\epsfig{file=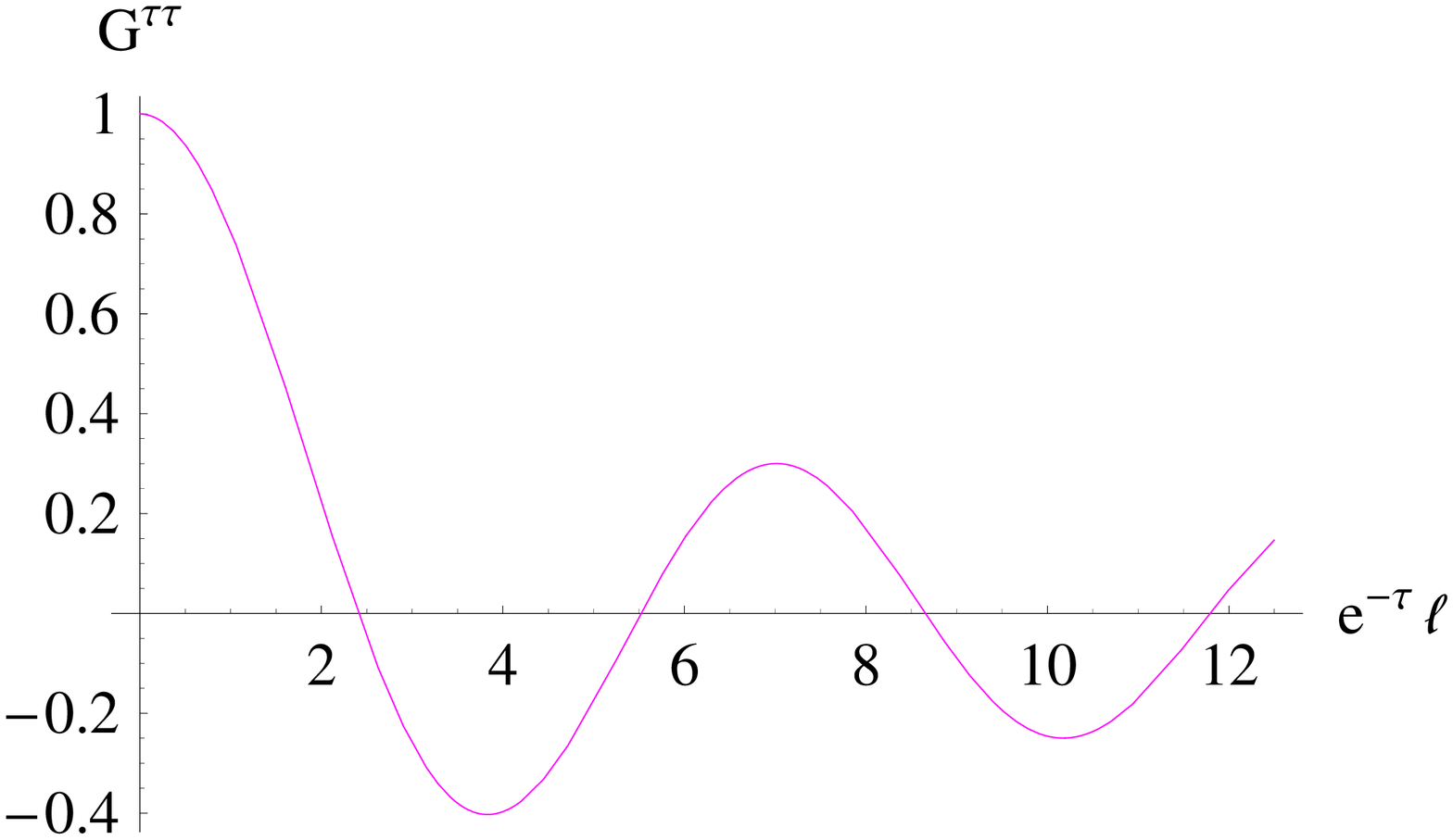,width=2.5in}
\end{center}
\caption{\footnotesize The leading behaviour of correlators
$G_{\t\t}$ (and $G_{\t\o}$), up to normalization
constants: as a function of $\t$ on the left with $\ell=1000$ and, on
the right, as a function of $\ell e^{-\t}$ }\label{1}
\end{figure}

\begin{figure}[h]
\renewcommand{\figurename}{Fig.}
\begin{center}
\epsfig{file=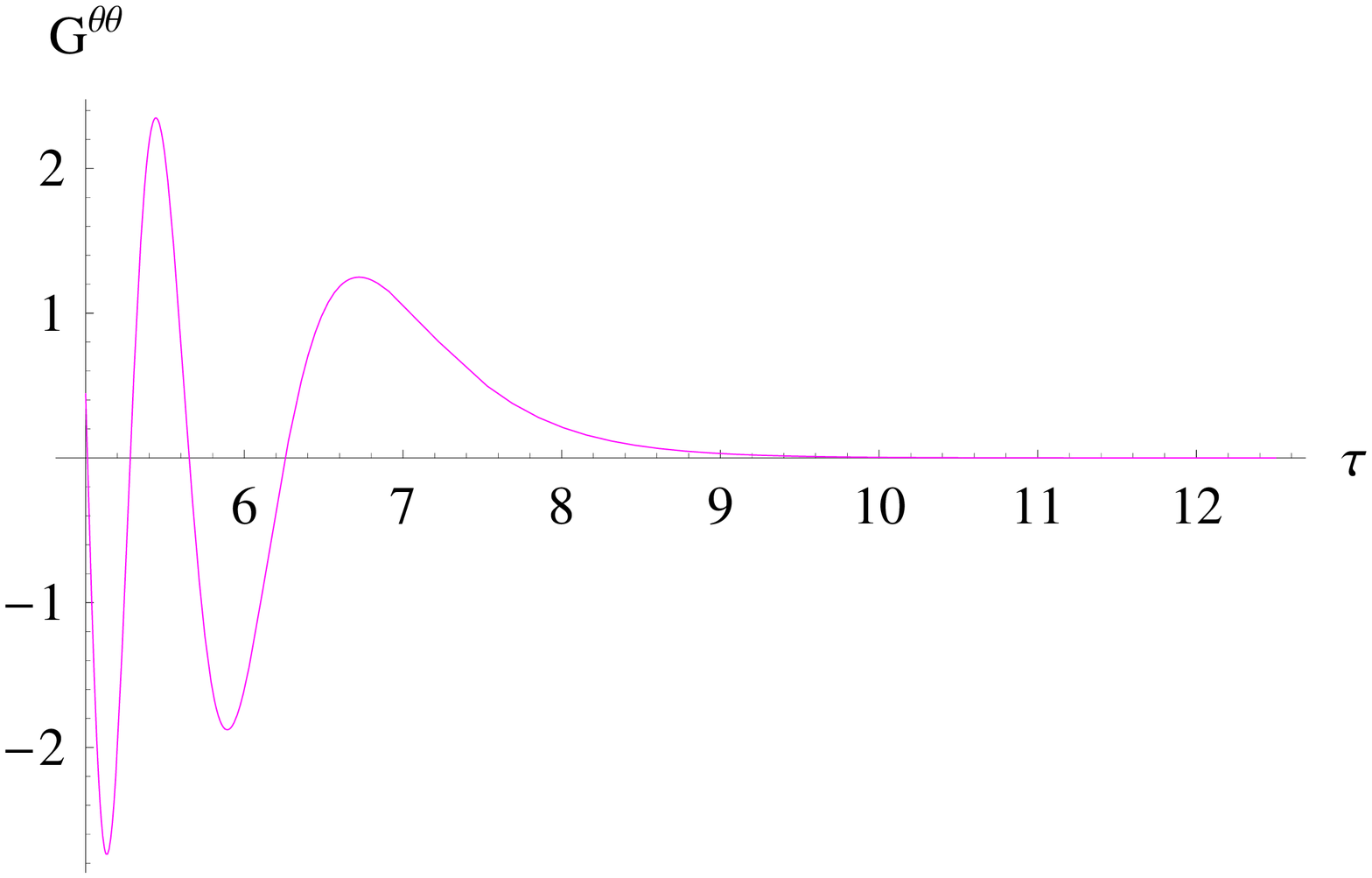,width=2.5in}\hspace{0.3in}
\epsfig{file=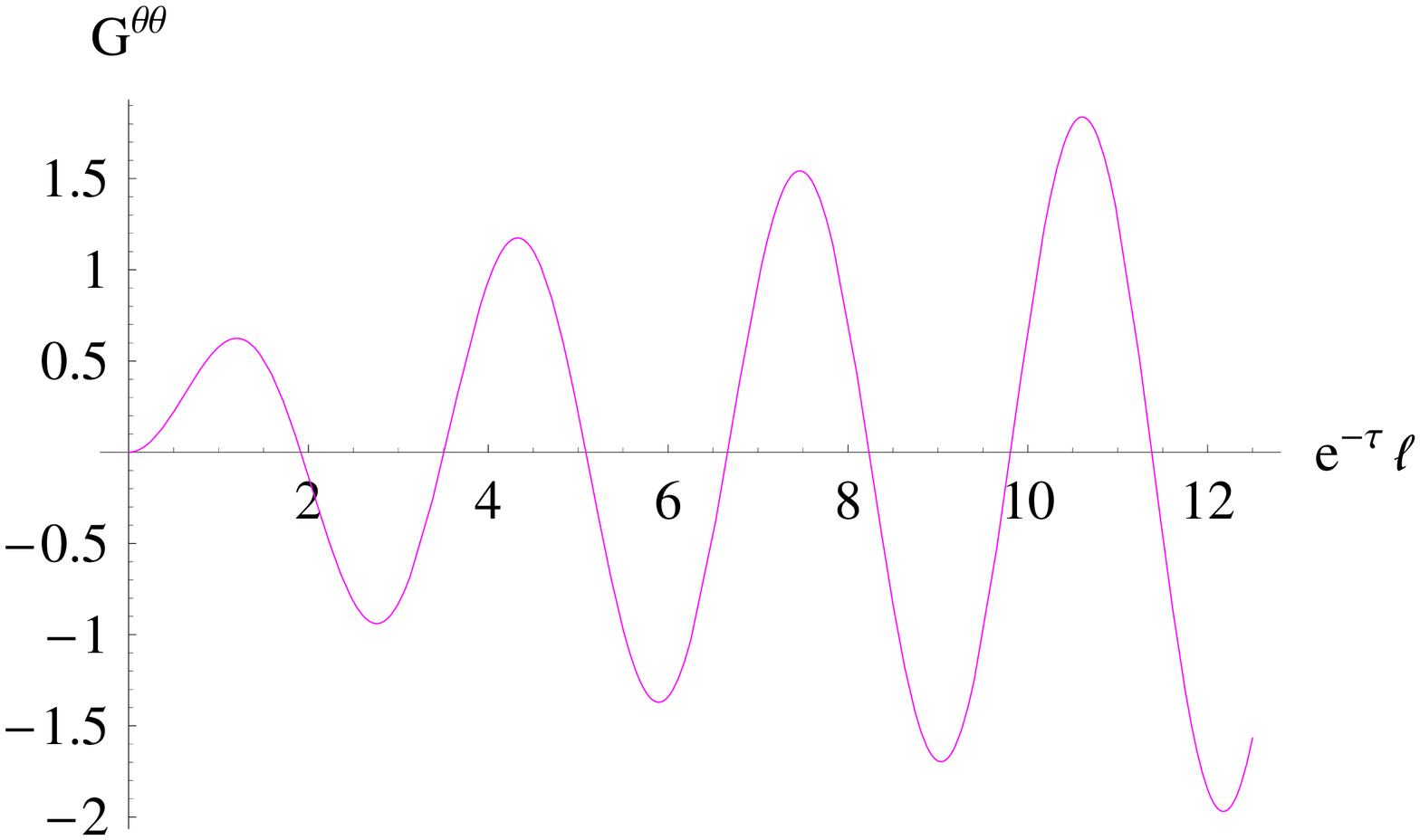,width=2.5in}
\end{center}
\caption{\footnotesize The leading behaviour of 
$G_{\o\o}$ (and $G_{\o\t}$): The left figure is plotted as a function
of time for $\ell=1000$, while the right hand side figure is a
function of $\ell e^{-\t}$.\label{2}}  
\end{figure}

For fixed physical wavelengths, $\ell\, e^{-\tau}$, or equivalently, at
the time when a harmonic $\ell$ crosses the horizon, the 
real time correlators are given by the exact results with the replacement 
\eqref{bessel}. The integral over $\nu$ can be easily evaluated using
the method of residues, and it turns out that the leading contribution
at late times is from the residue at $\nu=0$. Thus
\EQ
{
G^{\t\t}(\t,\,0;\ell),\,\,\, G^{\t\theta}(\t,\,0;\ell)\,\, 
\sim \,\,J_0(2\,\ell e^{-\tau} )
\label{late1}
}
and
\EQ
{
G^{\theta \t}(\t,0;\,\ell),\,\,\,G^{\theta\theta}(\t,0;\,\ell)\,\,\sim
\,\,\partial_\t J_0(2\,\ell e^{-\t}).
\label{late2}
}
%Therefore, at the leading order, the correlators $G_{\t\t}$ and
%$G_{\t\o}$ behave like ${}_0F_1 \left(;1+i \nu; -\tfrac{\ell^2
%    e^{-2\t}}{2}\right)$, while $G_{\o\t}$ and $G_{\o\o}$ behave like
%$\,_0F_1 \left(;1; \ell^2 e^{-2\t}\right)$. 
This  late time behaviour is depicted in
Figs. (\ref{1}) and (\ref{2}). 

%\begin{figure} 
%\renewcommand{\figurename}{Fig.} 
%\centering 
%\includegraphics[width=10cm]{Fig1.pdf} 
%\caption{The leading behaviour of correlators $G_{\t\t}$ and $G_{\t\o}$, up to some normalization constants.\label{1}} 
%\includegraphics[width=10cm]{Fig2.pdf} 
%\caption{The leading behaviour of $G_{\o\t}$ and $G_{\o\o}$.\label{2}} 
%\end{figure} 

The late time behaviour deduced above is not characteristic of
diffusion in de Sitter space. Suppose that the R-charge fluctuation
relaxed via diffusion modes, then the covariant conservation of the
R-current together with Fick's law would lead to the diffusion equation
\EQ
{
\partial_\t j^\tau  =  D \nabla_\theta\nabla^\theta \,j^\tau,
}
in $dS_3$, $D$ being the diffusion constant. 
The spherical harmonics of $j^\tau$ on the expanding spatial
spherical sections would then obey, 
\EQ
{
\partial_\tau j^\t_\ell = - D\,{\ell(\ell+1)\over \cosh^2\t}\, j^\t_\ell.
}
At late times $\tau \to \infty$ and large enough $\ell$, this is
solved by
\EQ
{
j^\t_\ell \sim \exp\left(\tfrac{1}{2} D\, \ell^2\,e^{-2\tau}\right).
\label{diffusive}
}

The large time behaviour of the Green's functions \eqref{late1} and
\eqref{late2} do not match up with expected diffusive relaxation
\eqref{diffusive} on $dS_3$. A natural reason for this is that the
rate of exponential expansion of the spatial section and the
Gibbons-Hawking temperature are both set by $R^{-1}$. Thus the mean
free path or the mean free time between collisions is comparable to
the expansion time scales so that the system never enters a
diffusive regime.

\section{The (small) AdS Bubble of Nothing}
The topological AdS black hole discussed above has a semiclassical
instability when
\EQ
{
r_\x < {R_{\rm AdS}\over {2\sqrt 2}}
}
which causes it to decay into a ``bubble of nothing'' in AdS
space. The instability only occurs if fermions have antiperiodic
boundary conditions in the $\chi$-direction. With periodic boundary
conditions for both bosons and fermions, the topological AdS black
hole is absolutely stable. 

As originally pointed out in \cite{Witten:1981gj} (and
\cite{Balasubramanian:2005bg} in the present context), the 
decay of a false vacuum in semiclassical gravity
is computed by the Euclidean bounce which has the same
asymptotics as the false vacuum in Euclidean signature. The bounce is
a solution to the Euclidean equations of motion with a non-conformal 
negative mode. In the context of the asymptotically (locally) AdS
spaces in question, the small Euclidean Schwarzschild solution
represents such a bounce solution. In Lorentzian signature, the
semiclassical picture of the decay process at time $t=0$ (say) involves
replacing the $t>0$ part of the false vacuum solution (the topological
black hole) with the appropriate analytic continuation 
of the Euclidean bounce to
Lorentzian signature. The analytic continuation of the small Euclidean
AdS black hole bounce which leads to 
$dS_3\times S^1$ boundary asymptotics, is the (small) AdS bubble-of-nothing.
\begin{figure}[h]
\renewcommand{\figurename}{Fig.}
\begin{center}
\epsfig{file=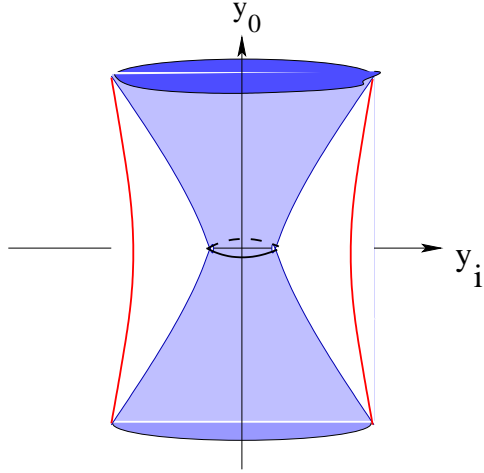,width=2.5in}
\end{center}
\caption{\footnotesize The global structure of the bubble
  geometry. The region inside the shaded region is empty, and its
  surface represents the de Sitter expansion of the bubble at $r=r_h$.
}\label{bubble}
\end{figure}
The metric for the AdS bubble-of-nothing solution is
\SP
{&ds^2=
  f(r)\;r_\x^2\;d\chi^2+f(r)^{-1}\;dr^2
+r^2\left(- {dt^2\over R_{\rm AdS^2}}+ 
\cosh^2\left({t\over R_{\rm AdS}}\right)d\Omega_2^2\right),\\ 
&f(r)=1+{r^2\over R_{\rm AdS}^2}-{r_h^2(R_{\rm AdS}^2+r_h^2)\over
  r^2}.
} 
In order to avoid a conical singularity in the interior, the
periodicity of the compact $\chi$ coordinate is related to $r_h$ as
\EQ
{
2\pi r_\x =2\pi R^2_{\rm AdS} \;{r_h\over r_h^2+ R_{\rm AdS}^2}.
} 
Passing to the dimensionless coordinates introduced earlier,
\EQ
{
\rho=\sqrt{(r/R_{\rm AdS})^2+1}, \qquad\tau={t\over R_{\rm
    AdS}},\qquad \tilde r_h = {r_h\over R_{\rm AdS}},
}
the metric becomes
\AL
{
&
ds^2= R^2_{\rm AdS}\left[\tilde f(\rho) \;{r_\x^2 \over R^2_{\rm AdS}}
\;d\chi^2 + \tilde f (\rho)^{-1}
\;{\rho^2\over \rho^2-1}\;d\rho^2+ (\rho^2-1)(-d\tau^2+\cosh^2\tau d\Omega_2^2)
\right]\\
& \tilde f(\rho)=\rho^2- {1\over \rho^2-1}\;{\tilde r_h^2}\;
\left(\tilde r_h^2+1\right).
}
In the AdS bubble of nothing 
spacetime, a slice of constant $\rho$ is $dS_3\times S^1$. 
The $S^1$, however, shrinks to zero size smoothly at $\rho=
\sqrt{\tilde r_h^2 +1}$. The shrinking circle is the cigar of the Euclidean
Schwarzschild solution. 
The boundary of the spacetime is approached as $\rho\rightarrow \infty$.
The semiclassical decay of the topological black hole at $\tau=0$,
results in the sudden appearance of a bubble of nothing in the
region of spacetime, $\rho^2 \leq 1+ \tilde r_h^2$.

\subsection{WKB for the AdS Bubble of Nothing}

An exact holographic computation of correlation functions in the AdS bubble of
nothing background appears difficult as analytical solutions to the
wave equation in this background are not known. Despite this, we may
obtain the boundary Green's function following a systematic
approximation. In particular we will employ the WKB approximation to
determine boundary correlation functions at high frequency and large mass.
The WKB approximation has been used successfully
\cite{Festuccia:2005pi}
to find high frequency Green's functions in the Big AdS-Schwarzschild
black hole. In this approximation, lines of isolated singularities (poles)
get replaced by branch cuts since, in the high frequency regime, the
separation between poles effectively goes to zero, as seen in our
example above in Section 2.3.1. 

We first take the massive scalar wave equation 
\EQ
{
{1\over \sqrt{-g}}\;\partial_\mu (g^{\mu\nu}\sqrt{- g}\;\partial_\nu \Phi)-
M^2 \Phi=0, 
}
in the bubble background 
and expand the scalar field in harmonics on
$dS_3\times S^1$ as in \eqref{decomp}. The harmonics $\Phi_n(\nu,
\rho )$ then satisfy a radial differential equation which can be put
in the form
\SP
{
&{d^2\Phi_n\over dx^2}+\left({1\over x-1}+{1\over x+\tilde r_h^2}+
{1\over x-\tilde r_h^2-1}\right){d\Phi_n\over dx}+
{1\over
4(x-1)(x+\tilde r_h^2)(x-\tilde r_h^2-1)}\times\\\\  
&\times\left({\nu^2+1}
-{n^2}{R^2_{\rm AdS}\over r_\x^2}{(x-1)^2
\over(x+\tilde r_h^2)(x-\tilde
r_h^2-1)} 
-{M^2R^2} (x-1)\right)\Phi_n=0,}
where $x=\rho^2$ and $n$ labels the momentum along the $S^1$. 
This is an ordinary differential
equation with four regular or nonessential 
singular points at $x=1, \,-\tilde r_h^2,\,
\tilde r_h^2+1$ and $\infty$. Analytical solutions for this type of
equation are unknown. In fact, a similar differential equation was
encountered in the computation of glueball masses at strong coupling 
in the three dimensional effective theory obtained from Euclidean
thermal ${\cal N}=4$ SYM \cite{Csaki:1998qr} (on ${\mathbb R^3}\times
S^1$ with SUSY-breaking boundary conditions). In that case the 
dual bulk geometry is the Euclidean black brane solution in
AdS space where the thermal circle shrinks to zero size smoothly.

The WKB solutions to the wave equation can be found after  
going to the Schr\"odinger form by introducing the variables
\AL
{
&\Phi={\Psi\over \sqrt{\rho(\rho^2-1)}},\\
&u\;=\;
{\tilde r_h\over 1+2\tilde r_h^2}\; 
\cot^{-1}\left({\rho\over \tilde
    r_h}\right)+{\sqrt{1+\tilde r_h^2}\over{1+2\tilde r_h^2}}\;\coth^{-1}
\left({\rho\over\sqrt {1+\tilde r_h^2}}\right).
%{du\over
%  d\rho}=-{\rho^2\over \rho^2(\rho^2-1)-\tilde r_h^2(\tilde r_h^2+1)}\;.
\label{newvar}
}
These are the natural generalizations of \eqref{rg1} and \eqref{rg2}
to the bubble of nothing geometry. The cigar in the geometry gets
smoothly capped off at $\rho=\sqrt{\tilde r_h^2+1}$, where the
spacetime ends. In terms of
the  $u$ coordinate, this occurs as  $u\rightarrow\infty$. 
In terms of the harmonics of $\Psi$ 
on the $dS^3\times S^1$ slices, as in \eqref{decomp}, we obtain the
Schr\"odinger equation 
obeyed by the harmonics $\Psi_n(\nu,u)$ in the (small) AdS
bubble of nothing:
\AL
{
&-{d^2\over du^2}\;\Psi_{n}(\nu,u) 
+ \tilde V_{n}(\nu, u)\;\Psi_n(\nu, u)= 0, 
\label{schr2}\\\nonumber\\ 
\label{schr3}
&\tilde V_{n}(\nu, u)= 
\left( (MR)^2- {\nu^2+1\over \rho^2-1}\right)
\left(\rho^2-1-{1\over \rho^2}\;\tilde r_h^2\;(\tilde r_h^2+1)\right)+
\\\nonumber
&\qquad\qquad\quad+ n^2 {R^2_{\rm AdS}\over r_\x^2}\;{\rho^2-1\over
  \rho^2}+ {1\over 
  4\rho^2}\;
(15\rho^4-10\rho^2-1).
}
Here $\rho$ is implicitly a function of $u$, determined by the solution to 
\eqref{newvar}.
In addition to the fact that the Schr\"odinger potential is far more
complicated than \eqref{schrodinger1}, one crucial difference to the
TBH case is that the potential cannot be defined independent of the
frequency itself. This is due to the term proportional to $\tilde r_h$
in \eqref{schr3}. This situation is also in contrast to the
case of the big AdS-Schwarzschild black hole in
\cite{Festuccia:2005pi}. Here, for a
given frequency (and mass), we need to find the ``zero energy'' eigenstate
of the Schr\"odinger problem  \eqref{schr2}.

\begin{figure}[h]
\begin{center}
\epsfig{file=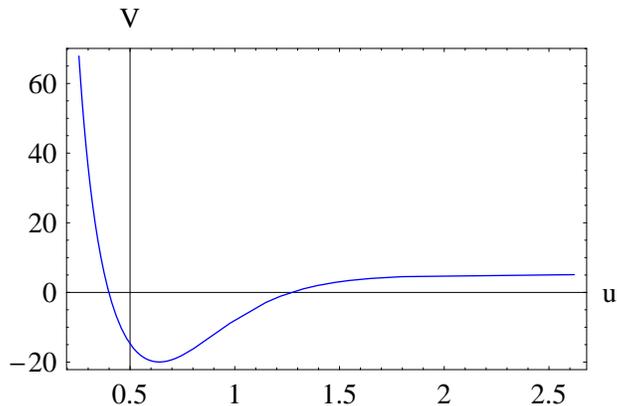,width=3.5in}
\end{center}
\caption{\footnotesize The Schr\"odinger potential for a
massive scalar in the AdS bubble of nothing. In the above plot the
dimensionless frequency $\nu=7$, the mass $MR= 2$ and $\tilde r_h = 
r_h/R_{\rm AdS} =1$.}
\label{schrpot1} 
\end{figure}

\begin{figure}[h]
\begin{center}
\epsfig{file=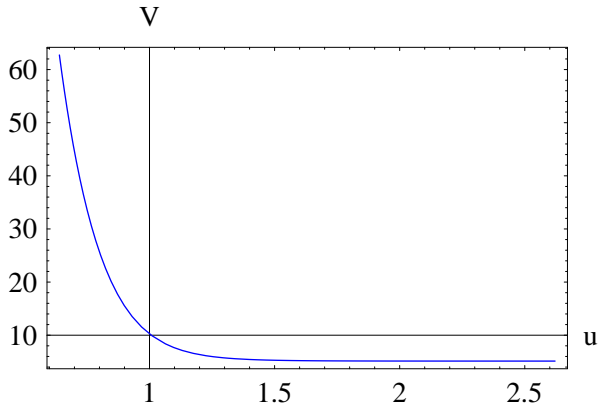,width=3.5in} 
\end{center}
\caption{\footnotesize The Schr\"odinger potential for 
$\nu=7$, the mass $MR= 7$ and $\tilde r_h = 
r_h/R_{\rm AdS} =1$. }
\label{schrpot2}
\end{figure}

Interestingly, 
the qualitative behaviour of the potential $\tilde V_n$ changes,
depending on the relative values of the mass and the frequency. This
is illustrated in Figures (\ref{schrpot1}) and (\ref{schrpot2}). 
The qualitative nature of the potential is easy to grasp in the high
frequency limit, wherein we take 
\EQ
{
\nu\rightarrow\infty,\quad MR\rightarrow\infty,\quad
\tilde \nu={\nu\over MR}\; {\rm fixed}.
}
For simplicity, we also set $n=0$, focussing attention on the
spatially homogeneous fields.
In this approximation,
\EQ
{
\tilde V_0(\nu, u)\rightarrow{\cal V}(\tilde\nu,u)=(MR)^2
\left(1-{\tilde\nu^2\over \rho^2-1}\right) 
\left(\rho^2-1 - {1\over \rho^2 }\;\tilde r_h^2(\tilde r_h^2+1)\right).
\label{wkbpot}
}
The potential is vanishing (having thrown away a subleading constant
in the large frequency limit) at the tip of the cigar $\rho =
\sqrt{\tilde r_h^2+1}$. Using $u\sim -\ln(\rho-\sqrt{\tilde
  r_h^2+1})$ near this point, it follows that  
${\cal V}(\tilde \nu,u)$ decays
exponentially as a function of $u$. The potential has another zero  
at $\rho=\sqrt{{\tilde\nu}^2+1}$. Now, since the spacetime ends at
$\rho=\sqrt{1+\tilde r_h^2}$, the number of zeroes of the potential 
depends on whether
$|\tilde\nu|$ is greater than or less than $\tilde r_h$. In particular, when
$|\tilde\nu| > \tilde r_h$, the potential energy has two zeroes or turning points, 
and qualitatively
resembles Fig. \ref{schrpot1}, while for $|\tilde\nu|< \tilde r_h$, it behaves 
as in Fig. (\ref{schrpot2}) (the potential asymptotes to a constant
different from zero in the figures; however this constant becomes
negligible in the high frequency limit).

Zero energy Schr\"odinger wave functions in the potential
\eqref{wkbpot} are marginally bound states for 
$|\tilde\nu| > \tilde r_h$, since the
potential has two turning points. For $|\tilde\nu| <\tilde r_h$, there is only
one turning point and the wave function should exhibit a qualitative
change in its behaviour at $|\tilde\nu|=\tilde r_h$. This also
strongly suggests that 
in the complex frequency plane, $\tilde\nu=\pm \tilde r_h$ should be
singularities of 
boundary correlation functions.

The zero energy wave function for the Schr\"odinger equation can be
obtained in the WKB approximation, where care needs to be taken in
applying the matching conditions at each of the turning points in the
potential. We treat the two cases $|\tilde\nu|> \tilde r_h$ and $|\tilde\nu|<
\tilde r_h$ separately.
\\\\
\underline{{\it WKB approximation for} $\tilde\nu^2 > \tilde r_h^2$}

There are two distinct regions in the potential: (i) the ``quantum
tunnelling region'', $\sqrt{1+\tilde\nu^2} < \rho <\infty$  which we label
as Region I, and (ii) the ``propagating region'', $\sqrt{1+\tilde
  r_h^2} < \rho < 
\sqrt{1+\tilde\nu^2}$ which we call Region II.

In Region I, we write the WKB solutions as
\EQ
{
\Psi_{\rm WKB}(\tilde\nu, u) = 
\;{1\over {\cal V}^{1/4}}\;\left(A_+
\exp\left({\int_{u_c}^u \sqrt {\cal V}
  \;du}\right) + A_-\;
\exp\left({-\int_{u_c}^u \sqrt {\cal V}
  \;du}\right)\right).
\label{wkb1}
}
where the classical turning point $u_c$ is defined by
\EQ
{
\rho(u_c) \equiv \sqrt{\tilde\nu^2+1}.
}
These represent the growing and decaying modes in the near boundary
region of the bulk geometry. This can be understood easily as
follows. 
%We find it useful to first define the WKB action
%\EQ{{\cal S}_I(u)=\int ^u \sqrt{\cal V}du.}
%For large $\rho$,
\EQ
{
{du\over d\rho}\bigg|_{\rho\to \infty} \approx - {1\over \rho^2}\;\;
\implies {\cal V} \approx (MR)^2\;\rho^2
}
which then immediately yields the near boundary WKB solution \eqref{wkb1}
for $\Psi$. 
This together with its relation \eqref{newvar}
to the massive bulk scalar field $\Phi$ implies
\EQ
{
\Phi_{\rm WKB}\big|_{\rm \rho\rightarrow\infty}\sim 
A_+\;(\cdots)\rho^{-2-MR}+A_- \;(\cdots)\;\rho^{-2+MR}
\label{asympwkb}
}
where the ellipses denote unspecified normalization constants.
The two power laws appearing in this solution are
precisely the 
normalizable and non-normalizable modes of the massive scalar
field, in the limit of large mass, 
in an asymptotically (locally) AdS spacetime. Following the
prescription for computing the retarded Green's functions, we
normalize $\Psi_{\rm WKB}$ \eqref{asympwkb} so that it approaches unity
near the boundary.

In the interior, however, for $\rho\leq
\sqrt{1+\tilde\nu^2}$ the solutions enter Region
II and become oscillatory. In Region II, we have
\EQ
{
\Psi_{\rm WKB}(\tilde\nu, u)=\;{1\over |{\cal V}|^{1/4}}\left(B_+\;
\exp\left(i{\int_{u_c}^u \sqrt {|{\cal V}|}
  \;d\rho}\right) + B_-\;
\exp\left({-i\int_{u_c}^u \sqrt {|{\cal V}|}
  \;du}\right)\right).
\label{wkb2}
}
The constants are uniquely determined by the WKB matching conditions at
the classical turning points of the potential ${\cal V}(\tilde\nu, u)$ and the
one normalization condition on $A_-$ near the boundary. The details of the WKB
matching conditions are explained in Appendix B. The crucial result
of the matching procedure is that the solution in the near boundary
region (Region I) has a normalizable mode with strength
\EQ
{
A_+ = - {1\over 2}A_-\;
\tan\left(\int_{\infty}^{u_c}\sqrt{|{\cal V}(\tilde\nu, u)|}du\right)
}
The argument in the above expession can, as
usual, be identified with the action of a zero energy classical particle 
trapped in the potential ${\cal V}(\tilde\nu,u)$ between
the two turning points $u_c$ and $u\rightarrow\infty$ (the latter
corresponding to $\rho\rightarrow\sqrt{1+\tilde r_h^2}$ where
spacetime ends).

The crucial difference between the bubble of nothing geometry and the
topological black hole is the boundary condition imposed on the
solutions to the wave equation in the interior. To extract retarded
correlators from the geometry with a horizon we impose an infalling
condition on the plane wave solutions near the horizon. 
In the bubble geometry, however, since spacetime
ends smoothly in the interior where the cigar caps off, there is no
freedom in choosing the boundary condition at the tip of the cigar --
we must require
regularity (normalizability) of solutions in the interior. This means
that the 
solution to the Schr\"odinger equation \eqref{schr2} must approach a constant 
exponentially as $u\to \infty$.
\begin{figure}[h]
\begin{center}
\epsfig{file=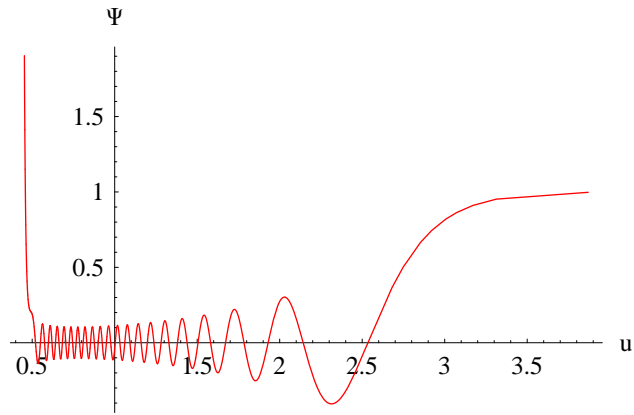,width=3.5in}
\end{center}
\caption{\footnotesize The exact numerical solution to the
Schr\"odinger problem for the AdS bubble of nothing. Here
$\nu R=200$, $MR=100$ and $r_h/R=1$. The solution approaches a
constant for $u\gg 1$, while the classically forbidden region is $0< u
< u_c \approx 0.49$.}
\label{wkbhi}
\end{figure}

The Green's function at high frequency is completely
determined ( up to contact terms) by the ratio $A_+/A_-$. After substituting the
solution \eqref{asympwkb} into the boundary action we find, 
\EQ
{
\tilde G_R(\tilde\nu) \approx \lim _{\epsilon\rightarrow 0}\;
{MR} \;\exp\left(2 \int_{u_c}^\epsilon \sqrt{{\cal V}(\tilde\nu,
    u)}\;du\right)  
\;\tan\left(\int_{\infty}^{u_c}
\sqrt{|{\cal V}(\tilde\nu, u)|} 
\;du\right).
\label{gfhi}
}
The exponential prefactor in this expression is the 
WKB transmission coefficient into the classically forbidden Region I.
The physically relevant contribution to the transmission coefficient
is the constant, $\epsilon$-independent term in an expansion around
the boundary $\epsilon \to 0$. The leading $\epsilon$ dependence, which
is an overall multiplicative constant proportional to
$\epsilon^{2MR}$, can be absorbed into the normalization of the
correlation function.

The first observation we can make without actually evaluating
\eqref{gfhi}, is that it has an infinite set of poles as a function of
$\tilde\nu$ on the real axis for $\tilde\nu^2> r_h^2$. These occur whenever
\EQ
{{\cal S}_{II}(\tilde\nu, \;MR)= \int_{\infty}^{u_c} du
\sqrt{|{\cal V}(\tilde\nu, u)|}=(n+ \tfrac{1}{2}){\pi}\,,
\quad n\in {\mathbb Z}.
}
That is, the semiclassical action in the propagating region is a
half-integral multiple of $\pi$. This is of course the condition for
the existence of a bound state wave function, and the poles in the
Green's function reflect the appearance of these bound states at
certain values of $\tilde\nu$ on the real axis. Recall that the
corresponding correlators in the topological black hole phase do not
have any poles on the real axis.

At first sight the poles on the real frequency axis might appear
somewhat surprising. However, they have the following natural physical
interpretation: the low energy physics of the gauge theory on the
boundary of the AdS bubble of nothing is that of  nonsupersymmetric
Yang-Mills theory 
at large $N$ (and strong 't Hooft coupling) on $dS_3$. The
antiperiodic boundary conditions on the spatial $S^1$ make all
fermionic excitations massive and the broken supersymmetry leads to
large radiative corrections to the scalar masses. In the strongly
coupled theory, the dynamical scale of the three dimensional effective
theory is expected to be set by $r_\x^{-1}$, the scale of the 
compact $S^1$ direction. When the 
Gibbons-Hawking temperature $T_H= 1/(2\pi
R)$ in $dS_3$ is smaller than $r_\x^{-1}$, we expect the gauge theory to be in a
confined phase where the degrees of freedom are gauge singlet
glueballs. The appearance of the isolated 
poles in the high frequency correlators is consistent with this
physical picture.

The WKB integral in Region II, 
in the high frequency approximation to the Green's function \eqref{gfhi}
can be expressed in terms
of complete elliptic integrals as
\SP
{&{\cal S}_{II}(\tilde\nu,\; MR)= 
\tfrac{1}{2}MR \int_{\tilde\nu^2+1}^{\tilde
    r_h^2+1}dx\;
{\sqrt{\tfrac{\tilde\nu^2}{x-1}-1}\over \sqrt{(x+\tilde
    r_h^2)(x-\tilde r_h^2-1)}}\\\\ 
&=i\,\tfrac{MR}{\sqrt{1+2\tilde r_h^2}}\,
\left[|\tilde\nu|\left(K(\tfrac{a}{b})-{\sqrt {\tfrac{b}{a}}}\;
K\left(\tfrac{b}{a}\right)\right)+
\tfrac{1}{|\tilde\nu|}(1+\tilde r_h^2)
\left(\Pi(a \big|\tfrac{a}{b})-{\sqrt {\tfrac{b}{a}}}
\;\Pi\left(b\big|{\tfrac{b}{a}}\right)\right)\right],\\\\
&%\qquad k= {a\over b}\,,\qquad 
a= \tfrac{\tilde\nu^2+1+\tilde r_h^2}{\tilde\nu^2}\;,\qquad 
b=\tfrac{1+ 2\tilde r_h^2}{\tilde r_h^2}\;, \qquad \tilde\nu^2 > r_h^2. 
}  
>From the general characteristics of these elliptic functions and their
singularities \cite{elliptic}, it  can be checked that ${\cal S}_{II}(\tilde\nu,
\;MR)$ has no singularities on the real axis for 
$\tilde\nu^2 > \tilde r_h^2$. Potential logarithmic branch points 
at $\tilde\nu^2=\tilde r_h^2$ and at $\tilde\nu=0$, cancel out
between the individual terms above. In fact, for any fixed value of
$\tilde r_h$, it  also follows that, for large $\tilde\nu$, the WKB integral
increases linearly with $\tilde\nu$.
\EQ
{
{\cal S}_{II} \propto |\tilde\nu|\,; \qquad |\tilde\nu|\gg 1.
} 
Hence for $|\tilde\nu|\gg \tilde r_h$, the propagator \eqref{gfhi} 
has approximately equally spaced simple poles on the real axis,
whenever ${\cal S}_{II}= (n+\tfrac{1}{2})\pi$. 

Although there are no other sources of singularities from $S_{II}$, 
the WKB transmission coefficient in Region I, which also enters the
Green's function \eqref{gfhi} can have branch point singularities on
the real axis,
\SP
{
&{\cal S}_I(\tilde\nu,\;MR)= - \tfrac{1}{2}MR
\int_{1+\tilde\nu^2}^{1/\epsilon^{2}}  
 \;dx\frac{\sqrt{1-\tfrac{\tilde\nu^2}{x-1}}}
{\sqrt{(x+\tilde r_h^2)(x-\tilde r_h^2-1)}}
\\\\
&= - \tfrac{MR}{\sqrt{1+2 \tilde r_h^2}}\;
\left(|\tilde\nu| \left(F\left(\csc^{-1}\sqrt{a},\;\tfrac{a}{b}\right)- 
K\left(\tfrac{a}{b}\right)\right)
-\tfrac{1}{|\tilde\nu| }(1+\tilde r_h^2)
\;\Pi\left(a|\tfrac{a}{b}\right)\right)\\
&\quad-\tfrac{1}{2}MR\left(\ln\left(\tfrac{2\epsilon^{-2}}{|\tilde\nu|
\sqrt{1+\tilde r_h^2}}\right) + i\pi\right).
}
This function is also free of any branch cuts at $\tilde\nu=\pm \tilde r_h$,
as can be checked by directly evaluating the integral at this
point. However, the logarithmic growth at large $\tilde\nu$ implies a branch
point at infinity.  

As an aside we mention that the high frequency limit in the
topological black hole phase \eqref{largemass} can be rederived by
formally setting $\tilde r_h=0$ in the WKB integrals and 
$\tilde G_R(\tilde \nu) \sim \exp( {2\,\cal S}_I)$.
\\\\
\underline{\it{WKB for $\tilde\nu^2< \tilde r_h^2$}}
\\\\
For low (real) frequencies $\tilde\nu^2 < \tilde r_h^2$, the nature of the 
WKB potential changes (Fig. (\ref{schrpot2})). Bound states are no
longer possible. The zero energy WKB solution is 
\EQ
{
\Psi_{\rm WKB}(\tilde\nu, u)= A_+ \;{1\over {\cal V}^{1/4}}\;\exp\left(
\int_{\infty}^u\sqrt{{\cal V}} \;du\right)\,+\,A_- {1\over {\cal V}^{1/4}}
\exp\left(
- \int_{\infty}^u\sqrt{{\cal V}} \, du\right).
} 
The relation between the two coefficients is determined by matching to
the wavefunction at large $u$, where the potential decays
exponentially and the wavefunction is a modified Bessel function (see
Appendix B). We find that
\EQ
{
A_+ = i A_-.
}
The aymptotics of this solution near the boundary at $u=\epsilon$
\eqref{asympwkb2}
allows to compute the boundary action and the Green's function
\EQ
{
\tilde G_R(\tilde\nu) \approx \lim_{\epsilon\to 0} 
i\,MR\,\exp({\cal S}_I ) = i \,MR\,
\exp\left(2 \int_\infty^\epsilon \sqrt{{\cal V}(\tilde\nu,u)}\, du\right).
}
\begin{figure}[h]
\begin{center}
\epsfig{file=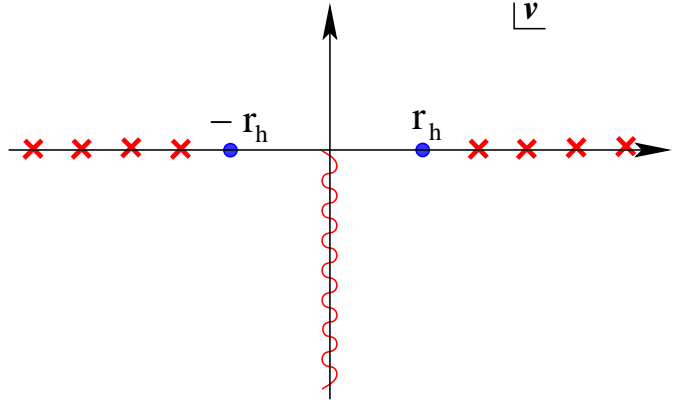,width=3.5in} 
\end{center}
\caption{\footnotesize The analytic structure of the boundary 
Green's function in the `` bubble of nothing'' phase in the WKB
approximation $\nu \to \infty$, $MR\to \infty$ with $\tilde\nu =
\nu/M R$ fixed. Approximately equally spaced simple poles on the
real  
axis for $\tilde\nu^2> \tilde r_h^2$, are accompanied by branch points at
$\tilde\nu=0$ and infinity.  
}
\label{wkbgf}
\end{figure}
In terms of elliptic functions the explicit form for the WKB action is
\SP
{
{\cal S}_I = &-\tfrac{MR}{\sqrt{1+2\tilde r_h^2}}\;
\left(|\tilde\nu| (F
\left(\csc^{-1}\sqrt{a},\;\tfrac{a}{b}\right)- 
\sqrt{\tfrac{b}{a}}K\left(\tfrac{b}{a}\right))
-\tfrac{1}{|\tilde\nu| }(1+\tilde
r_h^2)
\;\sqrt{\tfrac{b}{a}}\Pi\left(a\big |\tfrac{b}{a}\right)\right)\\
&-\tfrac{1}{2}MR\left(\ln\left(\tfrac{2\epsilon^{-2}}{|\tilde\nu|\,\sqrt{1+\tilde
        r_h^2}}\right)+ i \,\pi\right),
\label{wkbsing}
}
and the only singularity of this expression is the logarithmic singularity
at $\tilde\nu=0$ which complements the singularity at $\tilde\nu=\infty$ found
above.

The resulting analytic structure of the Green's function is summarized
in Fig. (\ref{wkbgf}), which is to be contrasted with corresponding
Green's function (at large mass and frequency) in the topological
black hole phase in Fig. (\ref{largetbh}).
As already dicussed earlier, the isolated poles on the
real axis indicate glueball states and 
that the gauge theory is in the confined phase,
wherein the radius of the spatial $S^1$ is much smaller than the $dS_3$
radius of curvature. This may be interpreted as a hadronized phase,
the de Sitter temperature being too low for the degrees of freedom to
be deconfined. In this context, we should point out that the high
frequency WKB analysis, where the frequencies are much larger than the
de Sitter cosmological constant, is basically a flat space limit and
thus the singularities of the Green's functions may be interpreted in
the standard way as in flat space. One feature of the propagator in
the bubble-of-nothing phase whose origin is not entirely clear is 
the branch point singularity at $\tilde\nu=0$. A similar branch point at
$\tilde \nu = -i$ was encountered in the high frequency limit in the
topological black hole phase. The associated branch cut was a
consequence of the apparent merger of the infinite set of quasinormal poles of
the topological black hole, in the high frequency limit. We do not
know of a similar interpretation of the branch cuts of \eqref{wkbsing}
and Fig. (\ref{wkbgf}).
\section{Summary and Discussion}
In this paper, we have studied real time correlators in strongly
coupled ${\cal N}=4$ SUSY Yang-Mills theory on a time-dependent background,
namely $dS_3\times S^1$. In particular, we have calculated the
retarded scalar glueball correlators and the R-charge current
correlators in the $\mathbb{Z}_N$-invariant phase.  

The retarded scalar glueball correlators have an infinite number of
poles in the lower half of the complex frequency plane, which
represent the topological black hole quasinormal frequencies. The
imaginary parts of these correlators are associated to the
Gibbons-Hawking temperature due to the cosmological horizon of
$dS_3$. These two facts suggest that the $\mathbb{Z}_N$ symmetric
phase of the boundary field theory corresponds to a deconfined plasma
in the exponentially expanding universe. 

We also computed the retarded correlators for the spatial spherical
harmonics of the conserved R-currents using the Son-Starinets
approach. Here we encountered a subtle point wherein we had to include
in our mode expansions, the effects of 
real, normalizable, discrete solutions to the de Sitter mode
equations, in order to obtained a retarded Green's function. The
corresponding 
frequency space correlator, appropriately defined in de Sitter space,
also has an infinite number of
poles in the lower half of the complex frequency plane, but these do
not appear to correspond to diffusive poles. The lack of hydrodynamic behavior
of the system is presumably due to the fact that the expansion rate of
$dS_3$ is of the same order as the Gibbons-Hawking temperature. Here,
we did not calculate the correlators of the stress-energy
tensor. However, using the same argument as above, we expect not to
find any hydrodynamic poles there either. 

In this paper, we have also calculated the retarded correlators of
scalar operators ${\cal O}_{\Delta}$, with conformal dimension $\Delta
\gg 1$, in both the $\mathbb{Z}_N$-invariant phase and $\mathbb{Z}_N$
broken bubble phase. Unlike the correlators of the $\mathbb{Z}_N$ symmetric
phase, the correlators in the $\mathbb{Z}_N$ broken phase feature an
infinite number of poles on the real frequency axis. These poles are
naturally associated
to bound glueball-like states, which suggests that this phase is a hadronized
phase, where the de Sitter temperature is too low to deconfine the
degrees of freedom (the Hubble parameter
is low compared to the dynamical scale of the effective, non-SUSY 3d theory
on $dS_3$). Since this geometry contains no horizon, Son-Starinets prescription \cite{Son:2002sd} is not applicable in this phase and we have restricted ourselves to the high frequency and large mass regime by using WKB approximation. Since the prescription proposed by Skenderis and van Rees \cite{Skenderis:2008dh,Skenderis:2008dg} does not rely on the existence of horizons in the geometry, it would be interesting to see whether one can obtain the retarded correlators beyond the high frequency limit using this prescription.

It is also interesting to note that the relevant boundary condition on the
horizon as prescribed by Son and Starinets \cite{Son:2002sd} implies
that the boundary theory is in the Euclidean or Bunch-Davies
vacuum. This is in agreement with Ref. \cite{Ross:2004cb}, where it
was argued that due to the fact that the $\alpha$-vacua Wightman
functions for the topological black hole develop singularities on the
event horizon, the preferred vacuum for the boundary field theory is
the Bunch-Davis vacuum. It would be interesting to see whether one can
understand the issue of $\alpha$-vacua ambiguity better by applying
the Skenderis-van Rees prescription \cite{Skenderis:2008dh,Skenderis:2008dg} to this set-up. 

\acknowledgments The authors would like to thank Carlos Hoyos,
Asad Naqvi, Simon Ross and Kostas Skenderis for useful discussions. 
J. H. is supported by the De Benedetti Family Fellowship in Physics
and in part by DOE grant DE-FG03-91-ER40682. J.R. is supported by an
STFC studentship.  
%where
%\EQ
%{
%S(u_c)= \int_{\infty}^{u_c}\sqrt{|\tilde V_0(\omega, u)|}du
%}

%\section{Small Bubble of Nothing}
%The small bubble of nothing in AdS space has the metric
%\EQ
%{ds^2={R^2\over z^2}\left(d\chi^2 f(z)+{dr^2\over f(z)} + (-dt^2+
%    \cosh^2 t \, d\Omega_2^2)\right)
%}
%where the $\chi$ coordinate is compact and yields a $U(1)$
%isometry. Here
%\EQ
%{
%f(z)=\left(1+z^2-{(z_+^2 +1)\over z_+^4}z^4\right).
%}
%Following Festuccia and Liu \cite{Festuccia:2005pi}, it is convenient
%to transform the radial coordinate to tortoise coordinate  
%\EQ
%{r \equiv \int_z^{z_+} \frac{z'^3}{f(z')} \, dz'.
%}
%The equation of motion for the spatially homogeneous mode then becomes
%\begin{eqnarray}
%\partial_r^2 \phi + F(r) \phi &=& 0; \\ 
%F(r) &=& \frac{f}{z^8} \left( z^2 \left(\nu^2 u^2 + 1\right) - \nu^2
%+ 4\right),  
%\end{eqnarray}
%where
%\EQ
%{\nu = \sqrt{4 + \left(mR\right)^2}; \,\,\,\,\,\, \omega \equiv \nu u.
%}
%If we write down $\phi = e^{\nu S}$, at the limit $\nu \rightarrow
%\infty$ we will get 
%\begin{eqnarray}
%\left(\partial_r S\right)^2 - V(r) &=& 0;\\
%V(r) &=& \frac{f}{z^8}\left(1 - u^2 z^2\right).
%\end{eqnarray}
%$z \in [0,1/u]$ is the classically forbidden region ({\bf Show graph
%  later}). Here, the solution can be written as 
%\EQ
%{\phi^{\rm WKB} (z) = \frac{e^{\nu {\cal Z}} \left(1 +
%  O\left(\nu^{-1}\right) \right)}{V(z)^{1\slash 4}}, 
%}
%with
%\EQ
%{{\cal Z} = - \int_{1/u}^z \frac{z'^3}{f(z')} \sqrt{V(z')} \, dz'.}
\newpage
\startappendix
\Appendix{Boundary action for bulk Maxwell fields}
\subsection{Nonvanishing $S^1$ momentum}
We provide here, the steps in the calculation leading to the boundary
action for fluctuations which are homogeneous on the spatial slices of
$dS_3$ but with momentum along the spatial $S^1$.
The equations of motion \eqref{gaugea},\eqref{gaugeb} can be solved
in terms of hypergeometric functions
\EQ
{
{\cal F}_n' = C_\t(\nu,\bar n)\;(1-z)^{-i(\nu-i)/2}\;{}_2F_1\left(
\tfrac{1}{2}+\tfrac{i}{2}(\bar n-\nu)\,,
\tfrac{1}{2}-\tfrac{i}{2}(\bar n+\nu)\,; 1-i\nu \,;1-z\right).
}
Here $C_\t$ is a frequency dependent constant, to be determined by
the boundary values of the fields.
Near the boundary of AdS space, the solution approaches
\SP
{
{\cal F}_n'\big|_{z\to 0} =&-\;C_\t\;\tfrac{\Gamma(1-i\nu)} 
{\Gamma\left(\tfrac{1}{2}+
\tfrac{i}{2}(\bar n-\nu)\right)
\Gamma\left(\tfrac{1}{2}-\tfrac{i}{2}(\bar n+\nu)\right)}\times \\
&\times \left(2
\gamma_E + \ln z +\psi\left(\tfrac{1}{2}+\tfrac{i}{2}(\bar
  n-\nu)\right) +
\psi\left(\tfrac{1}{2}-\tfrac{i}{2}(\bar n+\nu)\right)\right).
}
Imposing the boundary conditions at $z \to 0$
\EQ
{
\lim_{\e \to 0} {\cal F}_n(\nu,\epsilon)= {\cal F}^0_n(\nu), \qquad
\lim_{\e \to 0} {\cal G}_n(\nu,\epsilon)={\cal G}_n^0 (\nu),
}
from \eqref{eq3} we find that
\EQ
{C_\tau(\nu,\bar n) = - \tfrac 
{\Gamma\left(\tfrac{1}{2}+
\tfrac{i}{2}(\bar n-\nu)\right)
\Gamma\left(\tfrac{1}{2}-\tfrac{i}{2}(\bar
  n+\nu)\right)}{4\Gamma(1-i\nu)}\;\left(\bar n^2 {\cal F}_n^0(\nu) -
\tfrac{R_{\rm Ads}}
  {r_\x}\,\bar n (\nu-i) \,{\cal G}^0_n(\nu)
\right).
}
With the normalization fixed in terms of the boundary values of the
relevant fields we have that
\SP
{
{\cal F}_n'(\nu,\epsilon)=&\tfrac{1}{4}\left(\bar n^2 {\cal F}_n^0 -
  \tfrac{R_{\rm AdS}}{r_\x}\,\bar n (\nu-i) 
{\cal G}_n^0
\right)\;\times\\
&\left(2
\gamma_E + \ln \epsilon +\psi\left(\tfrac{1}{2}+\tfrac{i}{2}(\bar
  n-\nu)\right) + 
\psi\left(\tfrac{1}{2}-\tfrac{i}{2}(\bar n+\nu)\right)\right).
}
>From \eqref{cons} we also obtain
\EQ
{
{\cal G}_n'(\nu,\epsilon) = \tfrac{r_\x}{R_{\rm AdS}}{\nu+i\over \bar n}
{\cal F}_n'(\nu, \epsilon).
}
We can now plug these solutions into the boundary action to obtain the 
retarded R-current correlators. Following identical steps for the
normalizable (in time) modes, we have
\EQ
{
{\cal F}_n^{{\rm N}\prime}(\epsilon)=\tfrac{1}{4}\bar n^2{\cal
  F}_n^{{\rm N}0}\left(2
\gamma_E + \ln \epsilon +\psi\left(1+\tfrac{i}{2}\bar n\right) + 
\psi\left(1 -\tfrac{i}{2}\bar n \right)\right).
}

The induced boundary action for bulk Maxwell fields has the form
\SP
{S\big|_{z=\epsilon}&=
{4 \pi \over g^2_{SG}} \sum_{n} \frac{1}{2 \pi}\times \\
&\times\left(\int_{-\infty}^\infty 
{d\nu\over
  2\pi}\left( r_\x {\cal F}_{-n}^0(-\nu)\;{\cal
    F}_n'(\nu)\big|_{z=\epsilon} -  
\tfrac{R^2_{\rm AdS}}{r_\x}\, 
{\cal G}_{-n}^0(-\nu) \;{\cal G}_n'( \nu)\big|_{z=\epsilon}
\right) + r_\x{\cal F}_{-n}^{{\rm N}0}{\cal F}_{n}^{{\rm
    N}\prime}\big|_{z=\epsilon}\right), 
%\\
%& = \frac{N^2}{4 \pi R_{\rm AdS}} \sum_{n} \frac{1}{2\pi}\int {d\nu\over
%  2\pi}\left( r_\x {\cal F}_{-n}^0(-\nu)\;{\cal
%    F}_n'(\nu)\big|_{z=\epsilon} -  
%\tfrac{R^2_{\rm AdS}}{r_\x}\,{\cal G}_{-n}^0(-\nu) 
%\;{\cal G}_n'( \nu)\big|_{z=\epsilon}
%\right).
\label{bulkmaxwell1}
}
where $g^2_{SG}= 16\pi^2 R_{\rm AdS}/N^2$. The complete real time
retarded correlation functions for the R-currents, can now be accessed readily.
First we define the boundary values of our gauge fields as
\EQ
{
A_{\x,\t}(\epsilon,\t,\x)\equiv
\sum_n {e^{i n\x}\over 2\pi}A_{\x,\t}^n(\t)\,,
}
so that 
\EQ
{
{\cal G}_n^0(-\nu)=\int_{-\infty}^\infty\,A_\x^n(\t)\,{\cal T}_\x(\nu,\t)\cosh^2\t.
}
and similarly
\EQ
{
{\cal F}^0_n(-\nu)=\int_{-\infty}^\infty\,
A_\t^n(\t)\,{\cal T}_\t(\nu,\t)\cosh^2\t\,,\qquad
{\cal F}^{{\rm N}0}_n=\int_{-\infty}^\infty\,
A_\t^n(\t)\,{\cal T}_\t^{\rm N}(\t)\cosh^2\t\,.
}
Putting these ingredients together, we find that the boundary action
is 
\SP
{
&S\big|_{z=\epsilon}={N^2\over 4\pi R_{\rm
    AdS}}\;\sum_n \tfrac{1}{2\pi}\int_{-\infty}^\infty d\t\,\cosh^2\t
\int_{-\infty}^\infty d\t^\prime\,\cosh^2\t'\\
&\left[\int_{-\infty}^\infty{d\nu\over 2\pi} \left(2 \gamma_E+ \ln
    \epsilon + \psi\left(\tfrac{1}{2}+\tfrac{i}{2}(\bar n-\nu)\right) + 
\psi\left(\tfrac{1}{2} -\tfrac{i}{2}(\bar n+\nu) \right)\right)\times\right.\\
&\left.\times\left(
  \,A_\t^{-n}(\t)A_\t^{n}(\t')\,\tfrac{{\bar n}^2}{4}\,r_\x\,\,{\cal
    T}_\tau(\nu,\tau){\cal T}_\t(-\nu,\t')\,- \,A_\t^{-n}(\t)A_\x^n(\t')\,
{R_{\rm AdS}}
\tfrac{\bar n}{4}(\nu-i)\,\times \right.\right.\\\\
&\left.\left.\times {\cal T}_\t(\nu,\t){\cal T}_\x(-\nu,\t')\,-\, 
A_\x^{-n}(\t)A_\t^n(\t')\,R_{\rm AdS}\tfrac{\bar n}{4}(\nu+i)\,{\cal
  T}_\x(\nu,\t){\cal T}_\t(-\nu,\t')\,+\right.\right.\\\\
&\left.\left.+ A_\x^{-n}(\t)A_\x^{n}(\t')\tfrac{R^2_{\rm AdS}}{r_\x}
\tfrac{1}{4}(\nu^2+1){\cal T}_\x(\nu,\t){\cal T}_\x(-\nu,\t')\right)
+ r_\x A^{-n}_\t(\t)A^n_\t(\t')\times\right.\\\\
&\left.\times \tfrac{{\bar n}^2}{4}{\cal T}^{\rm N}(\t){\cal T}^{\rm N}(\t')
\left(2 \gamma_E+ \ln
    \epsilon + \psi\left(1+\tfrac{i}{2}\bar n\right)+ 
\psi\left(1 -\tfrac{i}{2}\bar n \right)\right)
\right].
\label{bulkmaxwell2}
}
\subsection{Nonzero momentum along the spatial slices of $dS_3$}
Below we fill in the steps in the derivation of the boundary action
for the Maxwell fields in the bulk.
The asymptotic form of the radial dependence of the bulk potential
${\cal F}'_\ell $, can be determined from \eqref{gaugesol}
\EQ
{
F_\ell\big|_{z\to 0} = C_\ell(\nu)\left(C_1(\nu) + C_2(\nu)\;\ln z
\right),
}
where
\AL
{
&C_1= - \frac{2\gamma_E
  +\psi(1-\tfrac{i\nu}{2})+\psi(-\tfrac{i\nu}{2})}{
\Gamma(1-\tfrac{i\nu}{2})\Gamma(-\tfrac{i\nu}{2})},\\
&C_2 = -{1\over \Gamma(1-\tfrac{i\nu}{2})\Gamma(-\tfrac{i\nu}{2})}.
}
We can solve for $C_\ell$ in terms of the boundary values of the gauge
potentials, using the bulk equation of motion \eqref{atau} for
${\cal F}_\ell'$ near the boundary which yields
\EQ
{
4 \int_{-\infty}^\infty {d\nu\over 2\pi}
\; \Gamma(1+i \nu)P_\ell^{-i\nu}(\tanh\tau)\;\ C_\ell(\nu) \;C_2(\nu)
= \ell(\ell+1)\left({\cal F}_\ell^0(\t)-\partial_\tau
  {\cal G}_\ell^0(\t)\right).
}
Note that the other equation of motion \eqref{atheta} for ${\cal G}_\ell'$,
just yields
the time derivative of this condition, so that we have only one
equation to determine the coefficient $C_\ell$. This equation can
be solved if we recall that the associated Legendre functions are
mutually orthogonal\footnote{The orthogonality of these functions for  
  purely imaginary order follows from the fact that they are
  eigenfunctions of the Schr\"odinger equation in the $\rm sech^2$
  potential, $\left(-\tfrac{d^2}{d\tau^2}- \ell(\ell+1)/\cosh^2\tau
  \right)P_\ell^{i\nu}(\tanh \tau)= \nu^2 P_\ell^{i\nu}(\tanh\tau)
$. In particular for $\nu\in {\mathbb R}$, these are scattering
states and are delta-function normalizable, and the eigenfunctions
corresponding to two different eigenvalues are orthogonal as usual.}
\EQ
{
\int_{-\infty}^\infty d\tau
\;P_\ell^{i\nu}(\tanh\tau)P_\ell^{-i\nu'}(\tanh\tau)=
{\delta(\nu-\nu') \over \Gamma(1-i\nu)\Gamma(1+i \nu)}.
}
Thus, we have
\EQ
{
C_\ell = {\ell(\ell+1)\over 4 C_2}\Gamma(1-i \nu)\int_{-\infty}
^\infty
d\tau'
\left({\cal F}_\ell^0(\t')-\partial_{\tau'} {\cal G}_\ell^0(\t')\right)
\; P_\ell^{i\nu}(\tanh\tau')
}
from which we obtain the solution 
\SP
{
&\cosh^2\tau\;{\cal F}'_\ell(\epsilon,\tau) = \; {\ell(\ell+1)\over 4}
\, 
\int_{-\infty}^\infty
d\tau'\left({\cal F}_\ell^0(\t')-\partial_{\tau'}
    {\cal G}_\ell^0(\t')\right)\\
&\bigg[\int_{-\infty}^\infty {d\nu\over 2\pi}\, {\pi \nu\over \sinh\pi\nu}\;
P_\ell^{-i\nu}(\tanh\tau)P_\ell^{i\nu}(\tanh\tau')\,\left(\ln \epsilon
  +2 \gamma_E+ 
\psi(-\tfrac{i\nu}{2})+\psi(1-\tfrac{i\nu}{2})\right) \\
& 
+ \sum_{m=1}^\ell m \tfrac{(\ell-m)!}{(\ell+m)!}\,
P_\ell^m(\tanh\t)P_\ell^m(\tanh\t')
\left(\ln\epsilon + 2\gamma_E+\psi\left( \tfrac{m}{2}\right)+
  \psi(1+\tfrac{m}{2})\right)\bigg]. \label{derivatives}
}
Plugging these back into the expression for the boundary action
\eqref{bdryaction}, we obtain the generating functional for two point
correlators of R-currents.
\Appendix{WKB matching conditions}
We explain below the matching conditions at the turning point(s) of the
WKB potential for the Schr\"odinger equation \eqref{schr2} in the AdS bubble of
nothing background.

\subsection{WKB matching conditions for $\nu > \tilde r_h$}

Near the turning point $u=u_c$ corresponding to $\rho = \sqrt{1+
  \nu^2}$, since we are well away from any extrema, we can assume
  that
\EQ
{
{\cal V}(\nu, u)=\kappa (u_c-u)+\ldots\,, \qquad {u \to u_c}.
}
In this region where the potential is basically linear, the exact
  solution in terms of Airy functions is
\EQ
{
\Psi\big|_{u\to u_c}=
A_+\; {2\sqrt\pi\over \kappa^{1/6}} \;{\rm Ai}\left(\kappa^{1/3}(u_c-u)\right)
+A_-\; {\sqrt\pi \over \kappa^{1/6}}\;{\rm Bi}\left(\kappa^{1/3}(u_c-u)\right)
.
\label{airy}}
The normalizations and constants of integration have been chosen
  carefully so that the exact solution near the
  turning point, in terms of Airy functions, matches the WKB solution
  \eqref{wkb1} 
  in Region I ($u< u_c$) away from the turning point. Now, we can
  continue the solution \eqref{airy} into Region II ($u> u_c$). For
  $u> u_c$, where the WKB solution \eqref{wkb2} should be valid, the
  Airy functions have the asymptotic form
\EQ
{
\Psi\big|_{u > u_c}\approx 
2 A_+\;{\sin\left(\tfrac{2}{3}\sqrt{\kappa}(u-u_c)^{3/2}+\tfrac{\pi}{4}\right)
\over \kappa^{1/4}(u-u_c)^{1/4}}+A_-\;
{\cos\left(\tfrac{2}{3}\sqrt{\kappa}(u-u_c)^{3/2}+\tfrac{\pi}{4}\right)
\over \kappa^{1/4}(u-u_c)^{1/4}}.
}
Comparison with \eqref{wkb2} near the turning point then implies
\EQ
{
B_+ = e^{i\tfrac{\pi}{4}}\left(\tfrac{1}{2}A_- - i A_+\right)\,;\quad
  B_-= e^{-i\tfrac{\pi}{4}}\left(\tfrac{1}{2}A_-+i A_+\right).
\label{cond1}}
There is yet another condition that emergies from the behaviour of the
  solutions near the second ``turning point'', $\rho\rightarrow\sqrt{1+\tilde
  r_h^2}$ or $u\rightarrow \infty$  
where the space ends. In this region we have
\EQ
{
{du\over d\rho}\big|_{\rho\to\sqrt{1+\tilde r_h^2}} \approx -
  {\sqrt{1+\tilde r_h^2}\over 2(2\tilde r_h^2+1)(\rho - \sqrt{1+\tilde r_h^2})}
}
so that
\EQ
{
\rho-\sqrt{1+\tilde r_h^2}\;\approx \;2\sqrt{1+\tilde r_h^2}
\exp\left(- 2 \tfrac{(1+ 2\tilde
  r_h^2)}{\sqrt{1+\tilde r_h^2}}\;u +\tfrac{\tilde r_h}{\sqrt{1+\tilde r_h^2}} \cot^{-1}
\left(\tfrac{\sqrt{1+\tilde r_h^2}}{\tilde r_h}\right)\right).
}
It follows then that, as a function of $u$, the high
frequency potential decays exponentially,
\EQ
{
{\cal V}(\nu, u)\big|_{u\rightarrow\infty}\approx (MR)^2
  \left(1-{\nu^2\over \tilde r_h^2}\right) 
%2 \frac{(1+ 2\tilde
%  r_h^2)}{\sqrt{1+\tilde r_h^2}} 
\;\exp\left(- 2 \tfrac{(1+ 2\tilde
  r_h^2)}{\sqrt{1+\tilde r_h^2}}\; u+{\rm constants}\right).
}
Note that for $\nu>\tilde r_h$, the potential approaches zero from
below. Let us define constants $A$ and $B$, in terms of which the potential
  is simply
\EQ
{
\tilde V_0(\nu, u)\big|_{u\rightarrow\infty}\approx - A e^{-B u},
}
where $A$ and $B$ can be read off easily from the expressions
  above. The Schr\"odinger equation with an exponentially decaying
  potential is solved exactly by Bessel functions:
\AL
{
&-\Psi''(u)-A e^{-B u}\Psi(u)=0,\\\nonumber\\
&\Psi= C_1 \;J_0\left(
2\tfrac{\sqrt{A}}{B} \,e^{-Bu/2}\right)+C_2\; Y_0\left(
2\tfrac{\sqrt{A}}{B} \,e^{-Bu/2}\right).
}
Recall that we are looking for a zero energy eigenfunction of the
  Schr\"odinger problem. This means that for a potential that
 vanishes at infinity, the corresponding (normalizable)
wavefunction can only be zero. This is an important difference to the
  black hole case where the wave functions are infalling plane waves
  at the horizon.
Requiring that the wave function $\Psi$ vanish or approach a constant
  as $u\to \infty$ then eliminates the term proportional to $Y_0$. Hence,
  in the exponentially decaying region 
\EQ
{
\Psi(u)\propto J_0\left(
2\tfrac{\sqrt{A}}{B} \,e^{-Bu/2}\right).
}
The WKB approximation should match onto the Bessel function for large
values of the argument of the Bessel function. Using the standard
asymptotic expansion for Bessel functions 
\EQ
{
J_0(x)\big|_{x\gg 1}\simeq {\cos(x)\over \sqrt{\pi x}}+
  {\sin(x)\over\sqrt{\pi x}}.
\label{asymj}
}
>From this we can deduce a relationship between the constants $B_+$ and
  $B_-$ in \eqref{wkb2}. To make this precise we define 
  the integral
\EQ
{{\cal S}_{II}(u) = \int^{u}_\infty \sqrt{|{\cal V}(\nu,u)|}\;du.
=- MR\int_{\tilde r_h}^{\rho^2-1} dx \sqrt{x^2-\nu^2\over{(x^2-\tilde r_h^2)
(x^2+1+\tilde r_h^2)}} 
}
where
\SP
{
&{\cal S}_{II}(u(\rho))= i\;\tfrac{MR}{\sqrt{1+2\tilde r_h^2}} 
\left[ \nu \left( F 
\left(\sin^{-1}\sqrt{\tfrac{1}{a}\tfrac{\rho^2+\tilde r_h^2}
          {\rho^2-1}},\;\;k\right)
-\tfrac{1}{\sqrt k} \;K\left(\tfrac{1}{k}\right)\right)+\right.
\\\\
&
\left.
\tfrac{1}{\nu}(1+\tilde r_h^2)
\left(\Pi\left(a;\,\,\sin^{-1} \sqrt{\tfrac{1}{a}
\tfrac{\rho^2+\tilde r_h^2}{\rho^2-1}} 
\,\bigg| \,k\right)-\tfrac{1}{\sqrt k}\;
\Pi\left(b\big|\tfrac{1}{k}\right)\right)\right],
\\\\
& a= \tfrac{\nu^2+1+\tilde r_h^2}{\nu^2}\;;\qquad 
b=\tfrac{1+2\tilde r_h^2}{\tilde r_h^2}\;;\qquad k= \tfrac{a}{b}.
}

Then the WKB solution in Region II is 
\EQ
{
\Psi_{\rm WKB}= {1\over |{\cal V}|^{1/4}}\left(B_+\;e^{i {\cal S}_{II}(u)-i
  {\cal S}_{II}(u_c)}+B_- e^{-i {\cal {S}}_{II}(u)+ i {\cal
    S}_{II}(u_c)}\right).
\label{wkblarge}
}
For large $u$, 
\EQ
{
{\cal S}_{II}(u)\big|_{u\gg 1} \approx \int_{\infty}^u \sqrt A e^{-B u/2}=
- 2{\sqrt A\over B}\; e^{-Bu/2}.
}
Using this result and comparing \eqref{wkblarge} to the asymptotics of
the Bessel function \eqref{asymj}, we find
\EQ
{
{B_+\over B_-}= \; i e^{2i\; {\cal S}_{II}(u_c)}.
\label{cond2}
} 
The final ingredient consists in determining $A_+$ and $A_-$. To this end
we first define 
\EQ
{{\cal S}_I(u)=\int^u\sqrt{{\cal V}(\nu,u)}\;du,}
which then gives us
\SP
{&{\cal S}_{I}(u(\rho))=\\ 
&\tfrac{MR}{\sqrt{1+2\tilde r_h^2}} \left[\nu 
F
\left(\sin^{-1}\left(\sqrt{\tfrac{1}{a}\tfrac{\rho^2+\tilde r_h^2}
          {\rho^2-1}}\right),\;k\right)+
\tfrac{1}{\nu}(1+\tilde r_h^2)\,\Pi\left(a;\sin^{-1}\left(\sqrt{
\tfrac{1}{a}\tfrac{\rho^2+\tilde r_h^2}
{\rho^2-1}}\right) \bigg | \,k\right)
\right],\\\\
& a= \tfrac{\nu^2+1+\tilde r_h^2}{\nu^2}\;;\qquad 
b=\tfrac{1+2\tilde r_h^2}{\tilde r_h^2}\;;\qquad k= \tfrac{a}{b}.
}  
Near the boundary $u\to 0$ or equivalently $\rho\to \infty$, we find
\SP
{
&\Psi_{\rm WKB}\approx \,A_-\tfrac{1}{\sqrt {MR}} \;\rho^{MR-\tfrac{1}{2}} \;e^{i\pi
  MR/2}\;\left({4\over 1+\tilde r_h^2}\right)^{MR/4}\;\nu^{-MR/2}\times\\
&\exp\left[\tfrac{MR}{\sqrt{1+2\tilde r_h^2}}\;
\left(\nu (F
\left(\csc^{-1}\sqrt{a},\;k\right)- K(k))-\tfrac{1}{\nu }(1+\tilde
r_h^2)
\;\Pi(a|k)\right)\right]\\
&+\; A_+\,\tfrac{1}{\sqrt {MR}} \;\rho^{-MR-\tfrac{1}{2}} \;e^{-i\pi
  MR/2}\;\left({4\over 1+\tilde r_h^2}\right)^{-MR/4}\nu^{MR/2}\times\\
&\exp\left[-\tfrac{MR}{\sqrt{1+2\tilde r_h^2}}\;
\left(\nu (F
\left(\csc^{-1}\sqrt{a},\;k\right)- K(k))-{1\over \nu }(1+\tilde
r_h^2)
\;\Pi(a|k)\right)\right].
}
Combining \eqref{cond1} and \eqref{cond2} we obtain
\EQ
{
A_+ = - A_-{1\over 2} \; \tan\left({\cal S}_{II}(u_c)\right).
}
%Finally we impose the normalization condition that the
%coefficient of the non-normalizable mode at the boundary should be set
%to unity so that
%\SP
%{
%&A_-= \sqrt{MR}\;e^{-i\pi MR/2} \left({4\over 1+\tilde
%    r_h^2}\right)^{-MR/4}\nu^{MR/2}\times\\ 
%&\exp\left[-{MR\over\sqrt{1+2\tilde r_h^2}}\;
%\left(\nu (F
%\left(\csc^{-1}\sqrt{a},\;k\right)- K(k))-{1\over \nu }(1+\tilde
%r_h^2)%
%\;\Pi(a|k)\right)\right]
%\label{aminus}
%}

\subsection{WKB matching for $|\nu|< \tilde r_h$}
When $|\nu|< \tilde r_h$, the potential energy ${\cal V}(\nu, u)$ is a
monotonic function of $u$ which exponentially vanishes as
$u\rightarrow\infty$. Now, the potential has effectively only one
turning point and the wave function has no region where it
propagates. The WKB solution \eqref{wkb1} in Region I should smoothly
match onto the exact solution of 
\EQ{-\psi''(u)+ A e^{-Bu}\psi(u)=0\,,\qquad A,B>0.}
The solutions to these are the modified Bessel's functions. Enforcing regular
behaviour as $u\to\infty$ picks out
\EQ
{
\psi(u)\propto I_{0}\left(2\tfrac{\sqrt A}{B}e^{-Bu/2}\right).
} 
The WKB approximation for $I_0(x)$ is valid when $x\gg 1$,
\EQ
{
I_0(x)\big|_{x \gg1}\simeq {1\over 2\sqrt {2\pi x}}\left(e^x+ i
  e^{-x}\right). 
}
We write the WKB solution to the wave equation in the bubble of
nothing background as
\EQ
{
\Psi_{\rm WKB}(\nu,u)= A_+ {1\over {\cal V}^{1/4}}\exp\left(
\int_{\infty}^u\sqrt{{\cal V}} du\right)+A_- {1\over {\cal V}^{1/4}}
\exp\left(
- \int_{\infty}^u\sqrt{{\cal V}} du\right).
}
Comparison with the modified Bessel function implies
\EQ
{
A_+=i A_-.
}
Near the boundary $u\to 0$, which is equivalent to $\rho\to \infty$,
we have
\SP
{
&\Psi_{\rm WKB}\approx 
\,A_-\tfrac{1}{\sqrt {MR}} \;\rho^{MR-\tfrac{1}{2}} \;e^{i\pi
  MR/2}\;\left({4\over 1+\tilde r_h^2}\right)^{MR/4}\;\nu^{-MR/2}\times\\
&\exp\left[\tfrac{MR}{\sqrt{1+2\tilde r_h^2}}\;
\left(\nu (F
\left(\csc^{-1}\sqrt{a},\;k\right)- 
\tfrac{1}{\sqrt k}\,K\left(\tfrac{1}{k}\right))
-\tfrac{1}{\nu}(1+\tilde
r_h^2)
\;\tfrac{1}{\sqrt k}\Pi\left(a\big |\tfrac{1}{k}\right)\right)\right]\\
&+\; A_+\tfrac{1}{\sqrt {MR}} \;\rho^{-MR-\tfrac{1}{2}} \;e^{-i\pi
  MR/2}\;\left({4\over 1+\tilde r_h^2}\right)^{-MR/4}\,\nu^{MR/2}\times\\
&\exp\left[-\tfrac{MR}{\sqrt{1+2\tilde r_h^2}}\;
\left(\nu (F
\left(\csc^{-1}\sqrt{a},\;k\right)- 
\tfrac{1}{\sqrt k}K\left(\tfrac{1}{k}\right))
-\tfrac{1}{\nu }(1+\tilde
r_h^2)
\;\tfrac{1}{\sqrt k}\Pi\left(a\big |\tfrac{1}{k}\right)\right)\right].
\label{asympwkb2}
}
%Normalizing the coefficent of the non-normalizable mode to unity, we get 
%\SP
%{
%&A_-=\sqrt{MR}e^{-i\pi MR/2}\left({4\over 1+\tilde
%r_h^2}\right)^{-MR/4}\times\\ 
%&\exp\left[-{MR\over\sqrt{1+2\tilde r_h^2}}\;
%\left(\nu (F
%\left(\csc^{-1}\sqrt{a},\;k\right)- {1\over \sqrt k}K\left({1\over k}\right))
%-{1\over \nu }(1+\tilde
%r_h^2)
%\;{1\over \sqrt k}\Pi\left(a\big |{1\over k}\right)\right)\right]
%}
%\AL
%{
%R_s(z)\approx & \; C\; z^{1+is}\; \Gamma[1-\sqrt{4+m^2}]\;{\Gamma(is)\over
%\Gamma({1\over2}(1+is-\sqrt{4+m^2}))^2}+{\rm c.c}\;+\\\nonumber
%& D\; z^{1+is}\; \Gamma[1+\sqrt{4+m^2}]\;{\Gamma(is)\over
%\Gamma({1\over2}(1+is+\sqrt{4+m^2}))^2}+{\rm c.c}
%}
%This fixes the ratio of $C$ and $D$. The solution near 
%the boundary $z\rightarrow 0$ is,
%\AL
%{
 %R_s(z)\rightarrow & \;C\;z^{2-\sqrt{4+m^2}}\left(1-{z^2\over
%    1-\sqrt{4+m^2}}{1\over 4}
%(1-\sqrt{4+m^2}-is)(1-\sqrt{4+m^2}+is)
%\right)\\\nonumber
%&\;D\;z^{2+\sqrt{4+m^2}}\left(1-{z^2\over
%    1+\sqrt{4+m^2}}{1\over 4}
%(1+\sqrt{4+m^2}-is)(1+\sqrt{4+m^2}+is)
%\right)
%}%%
%
%\subsection{Massless limit:}
%We now turn our attention to the wave equation for a masslees, minimally
%coupled scalar in the topological black hole geometry. The radial wave
%equation for a boundary s-wave is solved by the two independent
%functions
%
%The first of these has a non-normalizable boundary behaviour, 
%with the asymptotics:
%{\bf More here}%
%

\end{document}